\documentclass[acmlarge]{acmart}

\AtBeginDocument{%
  \providecommand\BibTeX{{%
    \normalfont B\kern-0.5em{\scshape i\kern-0.25em b}\kern-0.8em\TeX}}}

\setcopyright{acmcopyright}
\copyrightyear{2023}
\acmYear{2023}
\acmDOI{10.1145/1122445.1122456}

\acmJournal{IMWUT}
\acmVolume{37}
\acmNumber{4}
\acmArticle{52}
\acmMonth{1}
\usepackage{booktabs}
\usepackage{multirow}
\usepackage{graphicx}
\usepackage{tabularx}
\usepackage{placeins}
\usepackage{lscape}
\usepackage{ctable}
\usepackage{hhline}
\usepackage{mdframed}
\usepackage{tabularray}
\acmPrice{15.00}
\acmISBN{978-1-4503-XXXX-X/18/06}

\begin{document}

\title{Critiquing Self-report Practices for Human Mental and Wellbeing Computing at Ubicomp}
\author{Nan Gao}
\authornote{Both authors contributed equally to this research.}
\orcid{0000-0002-9694-2689}
\affiliation{%
  \institution{Department of Computer Science and Technology,}
  \institution{Tsinghua University}
  \city{Beijing}
  \country{China}
}
\affiliation{%
  \institution{University of New South Wales (UNSW)}
  \city{Sydney}
  \country{Australia}
  \postcode{1466}
}
\orcid{0000-0002-9694-2689}
\email{nangao@tsinghua.edu.cn}

\author{Soundariya Ananthan}
\authornotemark[1]
\affiliation{%
  \institution{University of New South Wales (UNSW)}
  \city{Sydney}
  \country{Australia}
  \postcode{1466}
}
\email{a.soundariya@gmail.com}

\author{Chun Yu}
\affiliation{%
  \institution{Department of Computer Science and Technology,}
  \institution{Tsinghua University}
  \city{Beijing}
  \country{China}
}
\email{chunyu@mail.tsinghua.edu.cn}

\author{Yuntao Wang}
\affiliation{%
  \institution{Department of Computer Science and Technology,}
  \institution{Tsinghua University}
  \city{Beijing}
  \country{China}
}
\email{yuntaowang@tsinghua.edu.cn}

\author{Flora D. Salim}
\email{flora.salim@unsw.edu.au}
\orcid{0000-0002-1237-1664}
\affiliation{%
  \institution{University of New South Wales (UNSW)}
  \city{Sydney}
  \country{Australia}
  \postcode{1466}
}
\email{flora.salim@unsw.edu.au}

\renewcommand{\shortauthors}{Gao et al.}

\begin{abstract}


Computing human mental and wellbeing is crucial to various domains, including health, education, and entertainment. However, the reliance on self-reporting in traditional research to establish ground truth often leads to methodological inconsistencies and susceptibility to response biases, thus hindering the effectiveness of modelling. This paper presents the first systematic methodological review of self-reporting practices in Ubicomp within the context of human mental and wellbeing computing. Drawing from existing survey research, we establish guidelines for self-reporting in human wellbeing studies and identify shortcomings in current practices at Ubicomp community. Furthermore, we explore the reliability of self-report as a means of ground truth and propose directions for improving ground truth measurement in this field. Ultimately, we emphasize the urgent need for methodological advancements to enhance human mental and wellbeing computing.

\end{abstract}





\maketitle

\section{Introduction}

Recently, Human-Centred Computing (HCC) \cite{jaimes2007guest} has gained immense importance due to its potential to enhance the interaction between humans and computers. It focuses on designing effective computer systems that take into account personal, social, and cultural factors, and addresses issues such as the relationships between computing technology and art, social, and cultural issues \cite{jaimes2007guest}. HCC has benefitted multiple fields such as Human-Computer Interaction (HCI) \cite{sinha2010human}, Computer-Supported Cooperative Work (CSCW) \cite{pratt2004incorporating}, User-Centred Design \cite{abras2004user}, Cognitive Psychology \cite{solso2005cognitive}, Sociology \cite{lupton2014self}, and Anthropology \cite{miller2020digital}, etc. The utilization of HCC technologies presents significant potential for enhancing human wellbeing through the development of early detection and intervention techniques for mental health. Such techniques, including emotion recognition \cite{dzedzickis2020human}, engagement detection \cite{gao2020n}, and interventions (e.g., cognitive training programs \cite{irazoki2020technologies}), can assist individuals in achieving and maintaining optimal emotional and mental states.

Despite several decades of research in HCC fields, those approaches have not yet been successful in transitioning from research trials to practical implementation in real-world scenarios, particularly when it comes to measuring human mental states and wellbeing such as emotion \cite{dzedzickis2020human}, depression \cite{wang2018trackingdepression}, engagement \cite{gao2020n}, anxiety \cite{Huang2016AssessingSA}, and more. In contrast, human physical activity recognition has demonstrated remarkable levels of accuracy, ranging from 83 to 100\% \cite{attal2015physical}, enabling its successful application in various real-world contexts, such as fitness trackers \cite{lockhart2012applications} and fall detection systems \cite{luvstrek2009fall}. Nevertheless, the assessment of mental wellbeing remains uniquely challenging due to its subjective nature, which encompasses emotions, thoughts, and subjective experiences that are inherently difficult to objectively quantify and measure. This difficulty in achieving objective measurements has resulted in relatively low accuracy, rendering it inadequate for real-world applications and effective interventions in practical settings. 


The primary reason for the low performance in computing mental wellbeing arises from the common practice of using self-report as the ground truth. This approach contrasts with the assessment of physical activity, which benefits from more objective measures such as direct observation and expert annotation. Mental wellbeing modelling, however, depends heavily on individual self-reports of experiences and emotions, introducing potential inaccuracies. For instance, Gao et al. \cite{gao2020n} employed physiological and environmental sensing to predict student engagement, using self-report data from the \textit{In-Class Student Engagement Questionnaires} (ISEQ) \cite{fuller2018development} as ground truth. Similarly, Wang et al. \cite{wang2018trackingdepression} tracked depression dynamics in college students using mobile phone and wearable sensing, with self-reported depression scores from the PHQ-8 \cite{kroenke2009phq8} and PHQ-4 \cite{kroenke2009phq4} questionnaires as the ground truth. While this reliance on self-reporting simplifies data collection, it introduces methodological inconsistencies (e.g., sample unrepresentativeness, threats to the reliability and validity of survey instruments) and susceptibility to various response biases (e,g., social desirability, extreme responding, and recall bias), which can significantly affect the effectiveness of mental wellbeing models.


%

The use of self-reporting in human mental computing research presents two primary concerns: the standardization of self-reporting practices themselves, and the choice of this research method. Challenges such as sample representativeness can distort data,  which in turn affects the effectiveness of modelling derived from this data. Factors like compensation schemes, non-response rates and withdrawal mechanisms also significantly impact the quality of self-report data \cite{kelley2003good}. Additionally, HCC studies, which differ from traditional survey research, require special considerations such as the integration of sensing data collection, the frequency of experience sampling methods (ESM), and the relation of these signals with psychological. Given these complexities, self-reporting, while seemingly straightforward, demands careful planning and meticulous execution in HCC research as the credibility of the self-report data is vital to prevent the risk of  \textit{`garbage in, garbage out'} \cite{kilkenny2018data}. 


While a few studies have explored the limitations of relying on self-report measures as ground truth for HCC \cite{gao2021investigating, semanticgap}, little progress has been made in developing effective solutions to address these issues. Gao et al. \cite{gao2021investigating} investigated the reliability of self-report and found that physiologically measured learning engagement and perceived engagement are not always consistent, underscoring the potential pitfalls of relying solely on subjective annotations as the basis for establishing ground truth. However, their study did not propose specific solutions to this problem. Das et al. \cite{semanticgap} found that the prediction performance of mental wellbeing depended on the method used to establish ground truth, with psychological-related features being more effective for self-report stress and behavioural-related signals being more effective for objective arousal signals (high arousal duration). While this approach represents an initial attempt to use alternative measures of ground truth, there are still concerns about whether arousal signals inferred from heart rate can be considered a reliable measure of mental wellbeing. 



In this work, we aim to raise awareness within the Ubicomp community, echoing standardising self-report practices and exploring reliable methods for establishing ground truth in HCC studies. To this end, we analyse 49 human mental computing studies in the \textit{ACM International Joint Conference on Pervasive and Ubiquitous Computing} (UbiComp) and the \textit{Proceedings of the ACM on Interactive, Mobile, Wearable and Ubiquitous Technologies} (IMWUT), and conducted a systematic literature review. We demonstrate comprehensive guidelines of self-reporting practices in human mental and wellbeing computing studies, and emphasize the need for methodological evolution in this field, advocating for a shift away from traditional self-reporting towards a more reliable and diverse method. For clarity, our study focuses on HCC studies related to human mental states and wellbeing computation, excluding physical behaviours computation, which is already well-established in the field. Specifically, our contributions are as follows:
\begin{itemize}

    \item Recognizing that self-reporting is the predominant method for measuring ground truth in HCC studies, we formulate a set of guidelines for self-reporting practices. It aims to enhance community standards and enhance the credibility of future research in the field of human wellbeing computing.
    \item Based on our analysis of 49 Ubicomp papers and the evaluation of self-reporting practices using the proposed guidelines, we identified substantial deficiencies and discrepancies in the self-report methodologies employed in current HCC studies.
    
    \item We discussed the reliability of self-report data as ground truth and demonstrated methodologies to enhance ground truth measurement in HCC studies. We also point out the future directions to improve human mental and wellbeing computing. 
\end{itemize}


The structure of the remaining paper is as follows. Section \ref{sec:related work} offers an overview of the background related to ground truth measures in HCC studies. Section \ref{sec:self-report practice} presents an examination of current self-report practices in HCC studies, highlighting common pitfalls within this domain. The reliability of self-report as a ground truth measure in human-centred computing is demonstrated in Section \ref{sec:reliabilty}. Section \ref{sec:future directions} delves into the discussion of future directions aimed at establishing ground truth in HCC studies. Section \ref{sec:future work} indicates future directions to advancing  HCC studies. Finally, Section \ref{sec:conclusion} provides the concluding remarks for the paper.

\section{Background}
\label{sec:related work}
In this section, we introduce the commonly used self-report measures in HCC, as well as the psychological constructs that are of primary interest to researchers and data commonly used in HCC research. This section aims to provide readers with a foundational understanding of these topics and their relevance to HCC research.

\subsection{Survey and Experience Sampling Method as Self-Report Measures}

Self-reporting involves individuals reporting their symptoms, behaviours, beliefs, or attitudes through tests, measures, or surveys \cite{paulhus2007self}. These self-reports are typically done on paper, electronically, or through interviews. Self-reporting is widely used in human-related studies, including psychological research and HCC, while in the latter, the survey and experience sampling methods are particularly popular. 

Surveys are often preferred due to their cost-effectiveness and ability to gather data from a large group of participants, covering a wide range of information such as basic demographics and specific social or behavioural factors \cite{Mutepfa2019}. Surveys can be conducted through questionnaires or interviews, but questionnaires are favoured for their ease of administration on a large scale. Typically, they can be categorized as either cross-sectional or longitudinal, with the latter involving data collection at multiple time points \cite{fox}). However, it should be noted that longitudinal surveys may not always be optimal due to their reliance on participants' cognitive abilities. 
To overcome this limitation, the \textit{Experience Sampling Method} (ESM) and \textit{Ecological Momentary Assessment} (EMA) \cite{bos2015experience} were proposed. These methods provide repeated snapshots of individuals' subjective information, thereby reducing reliance on memory and increasing response validity by gathering data at multiple points throughout the day or a specific period \cite{SONNENBERG20121037}.
Additionally, these methods allow for the storage of contextual details such as time and location, enabling the examination of temporal changes in participants' experiences and behaviours \cite{Trull2009UsingES}. It is worth noting that ESM and EMA are often used interchangeably in \cite{Wallace2018TheCS, van2017gamification}. In this paper, we refer to the term ESM.

\subsection{Psychological Constructs in HCC Literature}

Psychological constructs play a critical role in describing behaviour patterns and understanding natural phenomena. Researchers heavily rely on these constructs to explore human behaviour, emotions, and thoughts \cite{peterson_2022}. To avoid confusion, it is crucial to establish clear definitions and differentiate between similar constructs. For example, terms like affect, emotion, and mood may seem similar but have distinct meanings. It is noteworthy that certain constructs can be measured objectively. Sleep, for instance, can be measured using sleep sensors, Fitbit devices \cite{Xu2019LeveragingRB}, or even mobile phones \cite{WangFirstGen,Wang2017PredictingST, Wang2016CrossCheckTP}. However, our research will not focus on easily measurable constructs, as they have already been extensively studied. Instead, we will primarily examine constructs that are typically measured subjectively. Particularly, we have identified several commonly used psychological constructs in HCC studies. 

\textit{Anxiety, Stress and Panic}. \textit{Anxiety} and \textit{stress} are often used interchangeably, but there is a distinction between the two. Stress is a reaction to specific events or situations that can trigger emotional responses. On the other hand, anxiety is characterized by persistent worry, tension, and uncertainty \cite{Rehman2020DepressionAA}. It is an ongoing state of unease that can be difficult to control which may lead to various mental and health problems if not addressed. Similarly, \textit{anxiety} and \textit{panic} are similar but differ in their onset and symptoms. Panic attacks occur suddenly and are intense episodes of fear, often accompanied by physical symptoms such as a racing heart, shortness of breath, and chest pain, while anxiety develops gradually and allows for anticipation and planning \cite{Steimer2002TheBO}. Some HCC studies include detecting stress \cite{Hovsepian2015cStressTA, Mishra2018TheCF}, stress resilience \cite{Identifying}, social anxiety \cite{Huang2016AssessingSA} and panic attacks \cite{Towards}.

\textit{Engagement}.
Engagement is defined as a three-part classification that includes emotional, cognitive, and behavioural components. The emotional component refers to a positive state of mind and satisfaction. The cognitive component involves intellectual commitment, while the behavioural component encompasses effort and participation \cite{engagement1}. Engagement has been extensively studied in the field of HCC, such as student engagement \cite{gao2022individual,gao2020n, disalvo2022reading}, emotional engagement \cite{Lascio2018UnobtrusiveAO}, game engagement \cite{Huynh2018EngageMonME}, and social engagement \cite{Hernndez2014UsingEA}.

\textit{Depression}. Depression is a serious psychological condition that significantly impacts a person's wellbeing and overall functioning. It is a medical illness that can cause persistent feelings of sadness, hopelessness, and a loss of interest or pleasure in activities that were once enjoyed, leading to difficulties in daily life and a decreased quality of life \cite{depression2012depression}. Depression is a commonly studied psychological construct in HCC literature, with research focusing on topics such as depression during Covid-19 \cite{Tlachac,Tlachac2022}, trajectories of depression \cite{Canzian2015TrajectoriesOD}, etc.


\textit{Personality}.
Personality is a popular psycholgocial construct that refers to the consistent patterns of thoughts, feelings, and behaviours that make a person unique. It is often considered as a predictor of performance in studies and work \cite{article}.  The most widely accepted theory of personality structure is the Big-5 personality traits, also known as the \textit{Five-Factor Model} (FFM), suggesting five broad dimensions capture the major features of personality: extraversion, agreeableness, conscientiousness, neuroticism, and openness \cite{John1999TheBF}. Researchers in HCC have extensively studied Big-5 personality traits using various data sources, such as predicting personality using smartphones \cite{gao2019predicting, Wang2018SensingBC}, social media \cite{golbeck2011predicting}, online videos \cite{biel2012facetube}. 


\textit{Affect, Emotion and Mood}. These constructs, though share similarities, have 
distinct characteristics and serve different roles in understanding human behaviour and experiences \cite{ekkekakis2013measurement}. Affect is the immediate display of emotion, observable through physical expressions such as facial expressions, postures, and vocal tones. 
Emotions, on the other hand, is a complex internal experience involving subjective feelings, physiological changes (e.g., increased heart rate), and behavioural responses (e.g., smiling and frowning). It is typically triggered by specific events or stimuli. Mood, in contrast, is a longer-lasting emotional state that is not tied to a specific incident. It can persist for hours, days, or even longer and have a significant impact on an individual's wellbeing. Recent HCC studies for studying these constructs include: personalized mood \cite{Li2020ExtractionAI}, mood instability \cite{Saha2017InferringMI}, mood changes \cite{Lee2016OSNMT}, compound emotion \cite{Moodexplorer}, affect \cite{Zhang2018TeamSenseAP,Samyoun}, etc.

\textit{Cognitive Load and Mental Workload}.
Cognitive load refers to the extent of working memory resources used when a person engages in a task completion \cite{article2}. It embodies the mental effort necessary to learn new information or perform specific activities. Mental workload, on the other hand, is a more comprehensive term that encompasses the overall burden on the cognitive system, including working memory, attentional resources, and other cognitive functions. Related  HCC studies include cognition load \cite{Wilson2021} modelling, interruption management \cite{Goyal2017}, mental workload prediction \cite{kosch2018look}, etc.



\textit{Loneliness}. Loneliness is a complex emotion characterized by feeling isolated and alone, regardless of the number of social interactions. It is about the quality and meaning of social relationships rather than just the quantity. Loneliness can affect the mental state and cognitive ability by disturbing the processing of brain and increasing the risk of cardiovascular attacks \cite{loneliness}. It is also a commonly studied psychological constructs in HCC studies \cite{ Wang2015SmartGPAHS}.

 \textit{Flourishing}.
Flourishing is a multidimensional construct used in positive psychology to describe the optimal state of individuals characterized by good mental health,
extending beyond mere happiness or life satisfaction. It encapsulates a prosperous condition when people experience a sense of purpose, personal growth and the realization of their potential, leading to a profound sense of fulfilment and contentment \cite{Huppert2011FlourishingAE}. The measurement of flourishing scores has been extensively utilized in HCC studies such as \cite{ Wang2015SmartGPAHS,Saha2017InferringMI}


\textit{Fatigue}.
Fatigue stands as a prominent concern impacting both physical health and mental wellbeing, particularly in conditions like Multiple sclerosis (MS). MS is a neurological disorder that primarily affects young adults and has no known cure. Managing MS involves symptom control through support and therapeutic interventions. However, managing its symptoms is typically done through support and treatment \cite{Tong2019TrackingFA}. Among the various symptoms experienced by MS patients, detecting fatigue is particularly notable and have been extensively studies in HCC communities \cite{GuoMSLife,Tong2019TrackingFA}.

%





\subsection{Reliability and Validity of Self-Report}


The integrity of research findings in the field of HCC heavily relies on the chosen method for data collection. Among the various methods available, self-reporting is widely used for collecting data from human participants. However, self-report data face numerous challenges that can significantly impact the validity and reliability of the results. Validity refers to the degree to which a study accurately reflects or assesses the specific concept that the researcher intends to measure. It includes face validity (direct participant feedback), content validity (expert evaluation), and criterion validity (correlation with real-life constructs) \cite{Choi2004ACO}. On the other hand, reliability refers to the consistency of results obtained from an experiment, test, or any measuring procedure upon repeated trials \cite{bias}. It includes test-retest reliability (measuring stability over time), internal consistency (measuring agreement among questionnaire items), and inter-rater consistency (measuring agreement between observers). To mitigate the threats to validity and reliability, the common practice in HCC studies is to rely on established instruments like PHQ-8 \cite{kroenke2009phq8}, PHQ-4 \cite{kroenke2009phq4}, and GAD \cite{spitzer2006brief}) as the ground truth. However, common threats such as sampling bias, response bias and nonresponse bias should still be considered.

Especially, response bias refers to tendencies for participants to respond inaccurately or falsely to questions \cite{furnham1982good}. Common response biases include: 
(1) \textit{Recall bias}: Participants may inaccurately remember past events or recollect them wrongly. 
(2) \textit{Social desirability bias}: Participants may answer in a way that is favorable to others, leading to over-reporting or under-reporting. 
(3) \textit{Agreement bias}: Participants tend to select statements with positive implications or agree to statements. 
(4) \textit{Order effect bias}: Participant responses may vary based on the order of the questions. 
(5) \textit{Mood bias}: Participants' mental state can impact their answers, leading to changes based on their mood.
(6) \textit{Central tendency bias}: Some participants consistently choose responses in the middle of the scale, avoiding extreme agreement or disagreement.
(7) \textit{Demand characteristic bias}: Participants may alter their responses based on their understanding of the survey's purpose.
(8) \textit{Random response bias}: Participants may guess or choose random answers when they are unsure or do not understand the question.

\section{Self-report Practices for HCC Studies at Ubicomp}
\label{sec:self-report practice}

\subsection{Source selection}

We performed a comprehensive selection of papers from the \textit{ACM Digital Library}\footnote{See the ACM Digital Library at \url{https://dl.acm.org/}}. Our focus was on contributions from the Ubicomp conferences over the past decade (from Ubicomp '23 to Ubicomp '13), recognized as leading venues for research in ubiquitous computing and human mental wellbeing sensing. The sources for our collection were the \textit{Proceedings of the ACM on Interactive, Mobile, Wearable and Ubiquitous Technologies} and the \textit{Proceedings of the International Joint Conference on Pervasive and Ubiquitous Computing}. We selected the papers with the term \textit{human sensing}, \textit{physiological signals}, \textit{sensors}, \textit{mobile sensing}, \textit{wearable}, \textit{stress}, \textit{emotion}, \textit{depression}, \textit{mood}, \textit{affect}, \textit{prediction}, \textit{engagement}, \textit{cognitive load}, \textit{anxiety}, \textit{health}, \textit{wellbeing}, \textit{sensing}, \textit{behaviour}, and \textit{mental health}. Our search query, designed to capture this range of topics, is demonstrated below:

\framebox{

\parbox[t][3cm]{13.9cm}{

\addvspace{0.2cm} 

"query":    {Title, Keyword, Abstract: ("human sensing" or "physiological signals" or "sensors" or "mobile sensing" or "wearable" or "stress" or "emotion" or "depression" or "mood" or "affect" or "prediction" or "engagement" or "cognitive" or "anxiety" or "health" or "wellbeing" or "sensing" or "behaviour" or "mental health") AND AllField: (questionnaire or survey or "self-report")} "filter": {Conference Collections: UbiComp: Ubiquitous Computing} {E-Publication Date: (01/01/2013 TO 10/30/2023)}, {Published in: Proceedings of the ACM on Interactive, Mobile, Wearable and Ubiquitous Technologies}

} 
}
\\






Selection criteria beyond keyword relevance included:

\begin{enumerate}
    \item Emphasis on human-centered computing within the fields health, wellbeing, and affect prediction.
    \item Exclusion of workshop and poster papers.
    \item Utilization of wearable devices or smartphones for data acquisition.
    \item Publications dated from 2013 to 2023.
\end{enumerate}

\subsection{Screening Criteria}

An initial keyword search returned a total of 1,257 papers. After a preliminary review of titles and abstracts, 108 papers were identified that satisfied our selection criteria. Subsequent full-text review led to the exclusion of papers not providing adequate ground truth data collection methodologies. To analyze the information consistently, we tabulated critical data from each paper, such as questionnaires used, methodologies employed, types of devices utilized, and other pertinent details relevant to our study.



This process resulted in a refined collection of 49 papers. 
Table \ref{tab: papers_num_ground_truth} presents the reviewed Ubicomp papers and their respective types of ground truth. The predominant method for collecting ground truth is self-reporting, which was utilized in 93.87\% (46 out of 49) of the papers. This preference highlights the prevalence of self-report as the primary approach in human mental wellbeing computing within the Ubicomp field. Among the self-reporting instruments, 38 out of 46 papers relied on well-established surveys for data collection, such as the \textit{Perceived Stress Scale} (PSS), \textit{Big Five Inventory} (BFI), and \textit{Patient Health Questionnaire} (PHQ), among others. However, 8 out of 38 studies made modifications to these established surveys to better align them with the specific context of their research. Furthermore, a notable finding is that 8 papers developed custom-designed surveys tailored to their specific study goals. More details regarding these custom-designed surveys will be discussed in Section \ref{sec:self-report practice}. In contrast, the utilization of live observation as a method for collecting ground truth data was relatively limited, indicating a heavier reliance on slef-report data in assessing mental and wellbeing states within the Ubicomp research landscape.

\begin{table}
\small
\caption{List of Papers with Different Types of Ground Truth}
\label{tab: papers_num_ground_truth}
\begin{tabular}{p{4.2cm}p{1.8cm}p{7.5cm}}
\toprule
Types of Ground Truth                            & Paper Counts & Reference\\  \midrule
Established survey (unmodified)            &   30&\cite{Tlachac,Tlachac2022,Samyoun,GuoMSLife,Identifying,Wilson2021,levereging2021,Saha,Li2020ExtractionAI,Tong2019TrackingFA,Xu2019LeveragingRB,khwaja,Mirjafari2019DifferentiatingHA,Zhang2018TeamSenseAP,Wang2018SensingBC,Goyal2017,Wang2017PredictingST,Saha2017InferringMI,Costa2016EmotionCheckLB,Exler2016AWS,Lee2016OSNMT,Huang2016AssessingSA,Wang2015SmartGPAHS,Towards,Taylor2015UsingPS,studentlife,Generalization,GLOBEM,Antar}    \\\hline
Established survey (adapted)  & 8 & \cite{WangFirstGen,gao2020n,Huynh2018EngageMonME,Wang2018TrackingDD,Lascio2018UnobtrusiveAO,Gjoreski2016ContinuousSD,Canzian2015TrajectoriesOD, Hovsepian2015cStressTA} \\\hline
Custom-designed survey  & 8& \cite{Multi-Sensor,Nepal2020DetectingJP,Gashi2019UsingUW,Zhang2018MoodExplorerTC,Li2016EustressOD,Wang2016CrossCheckTP,Yu,Arkawa,microstress}      \\\hline
Direct observation                 & 3&\cite{Reading,Hernndez2014UsingEA,Mishra2018TheCF}       \\ 

\bottomrule
\end{tabular}%

\end{table}

\subsection{Coding Procedure}

Our analysis is based on the survey design principles in Arlene Fink's \textit{‘How to Conduct Surveys: A Step by Step Guide'} (Sixth Edition) \cite{fink2015conduct}. This comprehensive manual has become a vital resource for researchers as it provides guidance on various aspects of survey design, such as the selection of survey types, respondent inclusion criteria, survey frequency, and the intricacies of data analysis and result interpretation. The widespread acceptance and influence of this manual can be seen from its over 6000 citations to date. In addition to Fink's manual, we also referenced other leading publications in survey research, including \textit{‘Good Practice in the Conduct and Reporting of Survey Research'} by Kelley et al. \cite{kelley2003good} (more than 3200 citations to date),  \textit{‘The Survey Handbook'} by Fink \cite{fink2003survey} (more than 2900 citations to date), De et al. \cite{de2012international} (more than 1100 citations to date), enriching our procedural framework with recognized standards. 

As a result, we coded each paper in terms of the information reported about recruitment and participants. Besides, through examining the patterns and characteristics unique to human-centred computing studies, we assess the instruments, environments, sensing measures, data collection and post-processing methods for each paper. Table \ref{tab: codebook} describes the information we captured for each study.

\begin{table}[]
\footnotesize
\caption{Codebook used for the analysis of our sample}
\label{tab: codebook}
\begin{tabular}{@{}ll@{}}
\toprule
\textbf{Item}                        & \textbf{Description}                                                                                                                                                                           \\ \midrule
\textbf{Recruitment}                 &                                                                                                                                                                                                \\
1. Sample method                     & How were potential subjects identified? (e.g., random, systematic, convenience)                                                                                                                \\
2. Sample Size                       & How was the sample size decided?                                                                                                                                                               \\
3. Recruitment Method                & How, where, how many times, and by whom potential subjects were approached?                                                                                                                    \\
4. Participants approached           & How many participants were approached?                                                                                                                                                         \\
5. Participants agreed               & How many approached participants agree to participate?                                                                                                                                         \\
6. Non-response information          & How did those who agreed differ from those who did not agree with participating the study?                                                                                                     \\
                                     &                                                                                                                                                                                                \\
\textbf{Participants}                &                                                                                                                                                                                                \\
7. Age                               & Are the range, mean and STD of participants' age reported?                                                                                                                                     \\
8. Gender/Sex                        & Is the gender/sex of participants reported?                                                                                                                                                    \\
9. Ethnicity                         & Is the ethnicity of participants reported?                                                                                                                                                     \\
10. Occupation/job                   & Is the occupation/job of participants reported?                                                                                                                                                \\
11. Illness or health care           & Is the illness or health care information of participants reported?                                                                                                                            \\
12. Consent form                     & Did participants sign consent forms?                                                                                                                                                           \\
13. Compensation                     & Does the compensation mentioned?                                                                                                                                                               \\
14. Mechanism to leave study         & Were participants informed with the mechanism to leave study?                                                                                                                                  \\
                                     &                                                                                                                                                                                                \\
\textbf{Instruments and environment} &                                                                                                                                                                                                \\
15. Psychological constructs         & The psychologicl constructs studied in the research                                                                                                                                            \\
16. Existing/new instrument          & \begin{tabular}[c]{@{}l@{}}For new instruments, should provide a section outlining the steps taken to develop or test\\ the tool, including the results of psychological testing.\end{tabular} \\
17. Questionnaire/ESM                & Whether the instrument is a survey-based or ESM method                                                                                                                                         \\
18. Natural/lab settings             & Does the data collected in natural settings or lab settings?                                                                                                                                   \\
19. Sites                            & Does specific detail provided related to scenarios?                                                                                                                                            \\
                                     &                                                                                                                                                                                                \\
\textbf{Sensing Measures}            &                                                                                                                                                                                                \\
20. Device type                      & Types and details of the sensing device                                                                                                                                                        \\
21. Commercial/custom                & Whether the device is commerical or customed?                                                                                                                                                  \\
22. Sensing signals                  & Types of signals                                                                                                                                                                               \\
23. Relation to constructs           & Is the target psychological constructs measurable by the signals?                                                                                                                              \\
                                     &                                                                                                                                                                                                \\
\textbf{Data collection}             &                                                                                                                                                                                                \\
24. Survey administration            & How was the survey administrated (e.g., telephone, interview)                                                                                                                                  \\
25. Data collection duration         & Duration of the total data collection                                                                                                                                                          \\
26. Self-report frequency            & Frequency of self-report                                                                                                                                                                       \\
27. Self-report guidence             & Did the research guide participants to ensure effective self-report?                                                                                                                           \\
28. Response rate                    & What was the response rate?                                                                                                                                                                    \\
                                     &                                                                                                                                                                                                \\
\textbf{Post-processing}                &                                                                                                                                                                                                \\
29. Ground truth establishment*      & Method for estabilishing ground truth                                                                                                                                                          \\
30. Data quality ensurement*          & Provide specific details related to measures taken to improve data quality                                                                                                                     \\
31. Bias discussion*                  & Did researchers discuss about potential bias in the data collection?                                                                                                                           \\ \bottomrule
\end{tabular}
\begin{tabular}{@{}p{0.92\textwidth}}
\footnotesize 
Note: Asterisks (*) indicate the items that are not derived from existing literature
\end{tabular}
\end{table}


\subsection{Result}

In this subsection, we present the findings from our review, organized according to the sections of our codebook. We identify and discuss problematic issues while highlighting examples of best practices observed in the reviewed works.

\subsubsection{Recruitment}

In survey research, the selection of a representative sample from a well-defined sampling frame is critical for external validity, as it enables researchers to extrapolate findings to the broader population. Selecting who will be included in the sample requires careful consideration of various factors to achieve a comprehensive population profile \cite{kelley2003good}. However, our review reveals that none of the examined studies provided a full account of their participant recruitment process. While they disclosed sample sizes, all of them failed to perform essential sample size calculations, such as power analysis, which are fundamental for ensuring statistical robustness and the attainment of the study's objectives \cite{kang2021sample}.

A high rate of non-response can lead to misleading conclusions that may only reflect the views of the respondents, as indicated by Kelley \cite{kelley2003good}. French \cite{french1981methodological} found that non-respondents in patient satisfaction surveys are less likely to be satisfied than people who reply. It is critical to report the response rate and address the potential differences between respondents and non-respondents, with the implications of these differences. However, based on our review, none of Ubicomp papers report the non-response rate and information of differences, which raises concerns about data imbalance and the potential for skewed predictive models. For instance, Gao et al. \cite{gao2020n} developed a model to predict student engagement using physiological signals, yet the voluntary nature of student participation could mean those who opt-in are inherently more engaged, skewing results. Similarly, Wang et al. \cite{Wang2018SensingBC} employed mobile sensing data to predict the Big-5 personality traits; however, the likelihood that introverted individuals may opt out of participation could lead to an unbalanced model biased against introverted traits.



\begin{table}
\footnotesize
\caption{Overview of the self-report instruments in human-centred computing research}
\label{tab: instruments}
\begin{tabular}{lllll}
\toprule
\textbf{Category} & \textbf{Questionnaire} & \textbf{Variants} & \textbf{Papers} \\
\midrule
\textit{\multirow{3}{*}{Anxiety}} & State-Trait Anxiety Inventory (STAI) \cite{Spielberger1970ManualFT} & 40 items & \cite{Samyoun,Saha,Costa2016EmotionCheckLB,Gjoreski2016ContinuousSD} \\
\cmidrule(l){2-4}
& Generalized Anxiety Disorder (GAD) \cite{spitzer2006brief} & GAD-7 & \cite{Tlachac} \\
\cmidrule(l){2-4}
& Social Interaction Anxiety Scale (SIAS) \cite{MATTICK1998455} & 20 items & \cite{Huang2016AssessingSA} \\
\midrule
\textit{\multirow{2}{*}{Stress}} & Perceived Stress Score (PSS) \cite{Cohen1983AGM} & 14 items & \cite{Wang2015SmartGPAHS,Hovsepian2015cStressTA,Saha2017InferringMI} \\
\cmidrule(l){2-4}
& Trier Social Stress Test (TSST) \cite{Kirschbaum1993TheS} & Lab & \cite{Samyoun} \\
\midrule
\textit{\multirow{3}{*}{Engagement}} & In-class Student Engagement Questionnaires (ISEQ) \cite{fuller2018development} & 6 items & \cite{gao2020n} \\
\cmidrule(l){2-4}
& University Student Engagement Inventory (USEI) \cite{Marco2016UniversitySE} & 15 items & \cite{Lascio2018UnobtrusiveAO} \\
\cmidrule(l){2-4}
& Game Engagement Questionnaire (GEQ) \cite{BROCKMYER2009624} & 19 items & \cite{Huynh2018EngageMonME} \\
\midrule
\textit{\multirow{6}{*}{Depression}} &  \multirow{3}{*}{Patient Health Questionnaire (PHQ)  \cite{KROENKE2010345}} & PHQ-4 & PHQ-4: \cite{Wang2018TrackingDD,GLOBEM} \\
&& PHQ-8 & PHQ-8: \cite{Wang2018TrackingDD,Canzian2015TrajectoriesOD} \\
&& PHQ-9 & PHQ-9: \cite{Wang2015SmartGPAHS,studentlife,Tlachac,Tlachac2022,Identifying} \\
\cmidrule(l){2-4}
& Center for Epidemiological Studies Depression Scale (CES-D) \cite{Lewinsohn1997CenterFE} & 20 items & \cite{GuoMSLife} \\
\cmidrule(l){2-4}
& Beck Depression Inventory-II (BDI-II) \cite{Beck1996ComparisonOB} & 21 items & \cite{Xu2019LeveragingRB,levereging2021,GLOBEM} \\
\cmidrule(l){2-4}
& Depression Anxiety and Stress Scale (DASS) \cite{Gloster2008PsychometricPO} & 21 items & \cite{Saha2017InferringMI} \\
\midrule
\textit{\multirow{2}{*}{Personality}} & \multirow{2}{*}{Big Five Personality (BFI)  \cite{John1999TheBF}} & 44 items & BFI-44: \cite{Wang2015SmartGPAHS,Li2020ExtractionAI,Wang2018SensingBC,Generalization,khwaja} \\
&& 60 items & BFI-60: \cite{Saha} \\
\midrule
\textit{Affect} & Positive and Negative Affect (PANAS-X) \cite{Crawford2004ThePA} & 10 items & \cite{Wang2015SmartGPAHS,Saha,Zhang2018TeamSenseAP,Lee2016OSNMT,GLOBEM} \\
\cmidrule(l){1-4}
\textit{Mood} & Multidimensional Mood Questionnaire (MDMQ) & 24 items & \cite{Exler2016AWS} \\
\midrule
\textit{\multirow{2}{*}{Cognitive load}} & NASA TLX \cite{article3} & 6 items & \cite{Taylor2015UsingPS,Goyal2017,Wilson2021} \\
\cmidrule(l){2-4}
& Shipley Scales \cite{Carlozzi2011} & 40 items & \cite{Saha} \\
\midrule
\textit{Fatigue} & Fatigue Severity Scale (FSS) \cite{Krupp1989TheFS} & 9 items & \cite{Tong2019TrackingFA,GuoMSLife} \\
\midrule
\textit{Loneliness} & UCLA Loneliness Scale \cite{ucla} & 20 items & \cite{Wang2015SmartGPAHS} \\
\midrule
\textit{Flourishing} & Flourishing Scale \cite{articleflo} & 8 items & \cite{Wang2015SmartGPAHS,Saha2017InferringMI} \\
\midrule
\textit{\multirow{2}{*}{Panic}} & Panic Disorder Severity Scale (PDSS) \cite{SHEAR2001293} & 7 items & \cite{Towards} \\
\cmidrule(l){2-4}
& Diagnostic and Statistical Manual of Mental Disorders (DSM-IV) \cite{cooper2001diagnostic} & N/A & \cite{Towards} \\
\midrule
\textit{Functioning} & Quality of Life in Neurological Disorders (Neuro-QoL) \cite{Cella1860} & 8 items & \cite{Antar} \\
\bottomrule
\end{tabular}%
\end{table}
\subsubsection{Participants}


Previous studies have highlighted the importance of reporting detailed participant information such as age, gender, ethnicity, occupation, and health status in survey research  \cite{fink2003survey, de2012international,kelley2003good}. These demographic and personal characteristics can significantly influence self-report data, affecting its reliability and validity. For instance, age differences can impact cognitive responses, which in turn affect self-report outcomes \cite{andrews1986quality}. Similarly, gender differences in emotional processing and expression are well-documented \cite{fink2003survey}, suggesting that gender distribution in study samples should be carefully reported and considered.


None of the papers in our sample fully described their participant information. While all papers reported the number of participants, detailed age data (range, mean, and standard deviation) was often missing. Surprisingly, 17 papers did not report any age-related information, 26 omitted age ranges, and failed to provide mean or standard deviation values. Except for 12 papers, most reported participants’ gender distribution. However, only 12 papers addressed ethnicity. The inclusion of ethnicity is crucial, as cultural factors can influence perceived mental health and wellbeing perceptions in HCC studies. Occupation, another factor influencing mental health due to varying stress levels and work environment, was not mentioned in 8 papers.


The ethical aspects of research, particularly informed consent and the mechanism to leave the study, are vital for ensuring participant autonomy and ethical research conduct \cite{de2012international}. Our review found that 19 papers did not report information related to consent forms, and only 9 described the mechanisms for participants to withdraw from the study.

Compensation has significant impacts on the quality of self-report data. For example, the Netherlands Official Statistics used booklets with ten stamps as gifts, and the nonresponse rate fell significantly \cite{de2012international}. Conversely, Stone et al.  \cite{stone1991measuring} observed that offering a \$250 incentive led to poor data quality, attributing this to a participant pool driven primarily by financial gain rather than genuine interest. Based on our review, 24 out of 49 papers introduce compensation or incentives, which took various forms ranging from monetary payments, technology gadgets, vouchers, and non-monetary gifts.

\subsubsection{Instruments and environment}
In human mental and wellbeing computing studies, the identification of psychological constructs under investigation is crucial. Based on our review,  the most frequently studied constructs are depression (cited in 10 papers), stress (7 papers), mood (7 papers), and engagement (5 papers). 
38 studies leveraged established survey instruments. Among these,  8 papers adapted the surveys to suit their specific contexts. For instance, Gao et al. \cite{gao2020n} made slight modifications to the \textit{In-class Student Engagement Questionnaires (ISEQ)} \cite{fuller2018development}, originally designed for university lectures (e.g., \textit{`I feel discouraged when I worked on the activities in class'}), to the high school class context (e.g., \textit{`I feel discouraged when we worked on something'}). While such adaptations can enhance the contextual relevance of the surveys, they raise concerns about reliability and validity \cite{Choi2004ACO}. Modifications, even minor ones, could alter the construct being measured or affect the instrument's psychometric properties. According to De et al. \cite{de2012international}, regardless of the form or the degree of change, it is wise to consider adapted questions as new questions and to test them accordingly. However, none of the papers that utilized adopted established surveys examined the effects of these adaptations. 

8 papers utilized custom-designed self-report tools, often to ask quick questions using the ESM method. According to Kelley et al. \cite{kelley2003good}, \textit{`if a new survey tool is used, an entire section should be used to describe the steps undertaken to develop and test the tool, including psychometric assessment results'}. Unfortunately, in the 8 papers analyzed, custom-designed self-report tools were used without any test of reliability and validity. Recognizing the significance of employing well-established survey instruments in HCC studies, we have compiled a list of commonly utilized self-report instruments in Table \ref{tab: instruments}, for reference and use in future research.





\subsubsection{Sensing Measures}
Unlike traditional survey research, HCC studies predominantly rely on self-report data as the ground truth, complemented by sensing data as predictive indicators. The initial step involves confirming the psychological constructs to be analyzed. Subsequently, it's crucial to determine the types of signals to be collected, methodologies for data acquisition, and strategies to assure signal integrity. A critical aspect to be addressed is the feasibility of measuring targeted psychological constructs using these signals. The selection of appropriate signals is often contingent on the nature of the psychological constructs and the signal types. Broadly categorizing, these signals include physiological data from wearable devices, mobile sensing data from smartphones or desktops, environmental sensors, or a combination. Table \ref{tab:device} shows an overview of the devices used in HCC studies at Ubicomp.

\begin{table}[]
\footnotesize
\caption{Overview of the devices used in HCC studies at Ubicomp}
\label{tab:device}
\resizebox{\columnwidth}{!}{%
\renewcommand{\arraystretch}{1.2}
\begin{tabular}{>{\itshape}p{3.1cm}p{1.5cm}p{5.45cm}p{0.5cm}p{3cm}}
\toprule
\textbf{\text{Device}} & \textbf{Category}             & \textbf{Parameters}                                                                                                                                        & \textbf{Num.} & \textbf{Ref.}                                                                                                                                                                                                   \\ \midrule
Smartphone                       & Smartphone                     & Physical activity, sleep, app usage                                                                                                                        & 23                                         & \cite{WangFirstGen,Tlachac,Tlachac2022,Xu2019LeveragingRB,Mirjafari2019DifferentiatingHA,Wang2018TrackingDD,Wang2017PredictingST,Saha2017InferringMI,Wang2018SensingBC,Zhang2018MoodExplorerTC,levereging2021,khwaja,Exler2016AWS,Lee2016OSNMT,Li2016EustressOD,Wang2016CrossCheckTP,Huang2016AssessingSA,Wang2015SmartGPAHS,Canzian2015TrajectoriesOD,studentlife,Generalization,GLOBEM,Yu}                       \\  \hline
Empatica E4                      & Wristband                        & BVP, EDA, HR, ST                                                                                                                                      & 9                                          & \cite{Samyoun,Reading,gao2020n,Gashi2019UsingUW,Huynh2018EngageMonME,Lascio2018UnobtrusiveAO,Gjoreski2016ContinuousSD,Multi-Sensor,Wilson2021}                                                                        \\ \hline
Garmin Vivosmart                 & Smartwatch                       & Sleep, oxygen level, HR, breathing, physical activity                                                                                                                 & 3                                          & \cite{Saha,Nepal2020DetectingJP,Mirjafari2019DifferentiatingHA}                                                                                                                                                       \\ \hline
Polar H7                         & Chest strap                      & Activity speed, distance, HR                                                                                                                       & 4                                          & \cite{Mishra2018TheCF,Costa2016EmotionCheckLB,Li2016EustressOD,microstress}                                                                                                                                                       \\ \hline
Q sensor                         & Wrist sensor                     & EDA, ST, actigraphy                                                                                                                          & 2                                          & \cite{Hernndez2014UsingEA,Goyal2017}                                                                                                                                                                                   \\ \hline
Fitbit                           & Smartwatch                       & Sleep, physical activity                                                                                                                                   & 2                                          & \cite{levereging2021,GLOBEM}                                                                                                                                                                                          \\ \hline
Fitbit Flex 2                    & Smartwatch                       & Activity, calories burned, distance, sleep, steps                                                                                                          & 1                                          & \cite{Xu2019LeveragingRB}                                                                                                                                                                                             \\ \hline
Fitbit Charge 2                  & Smartwatch                       & Activity, calories burned, distance, HR, sleep, steps                                                                                              & 1                                          & \cite{Identifying}                                                                                                                                                                                                    \\ \hline
GENEActiv                        & Smartwatch                       & Physical activity, sleep, everyday behaviour                                                                                                                & 1                                          & \cite{GuoMSLife}                                                                                                                                                                                                      \\ \hline
Withings Aura                    & Bedside unit                     & Sleep, HR                                                                                                                                          & 1                                          & \cite{Tong2019TrackingFA}                                                                                                                                                                                             \\ \hline
Withings Activite Steel          & Smartwatch                       & Activity, calories burned, distance, sleep, steps                                                                                                          & 1                                          & \cite{Tong2019TrackingFA}                                                                                                                                                                                             \\ \hline
Withings Body Cardio             & Smartscale                       & Body composition, HR                                                                                                                               & 1                                          & \cite{Tong2019TrackingFA}                                                                                                                                                                                             \\ \hline
Sociometric badge                & Badge                            & Face-to-face interaction, conversation, physical proximity, physical activity                                         & 1                                          & \cite{Zhang2018TeamSenseAP}                                                                                                                                                                                           \\ \hline
Samsung Galaxy Tab S2            & Tablet                           & Physical activity, sleep, app usage                                                                                                                        & 1                                          & \cite{Huynh2018EngageMonME}                                                                                                                                                                                           \\ \hline
Microsoft Kinect                 & Add-on device                    & Postures and body movement                                                                                                                                 & 1                                          & \cite{Huynh2018EngageMonME}                                                                                                                                                                                           \\ \hline
Microsoft Band 2                 & Smartwatch                       & HR, EDA, ST, physical activity                                                                                                       & 1                                          & \cite{Wang2018TrackingDD}                                                                                                                                                                                             \\ \hline
Moto 360                         & Smartwatch                       & HR, step count                                                                                                                                     & 1                                          & \cite{Exler2016AWS}                                                                                                                                                                                                   \\ \hline
ekgMove                          & Chest belt                       & ECG, HR, HRV, steps, activity, and energy expenditure                                                                                              & 1                                          & \cite{Exler2016AWS}                                                                                                                                                                                                   \\ \hline
Zephyr BioPatchTM                & Body wear                        & HR, respiratory rate, physical activity                                                                                                            & 1                                          & \cite{Towards}                                                                                                                                                                                                        \\ \hline
AutoSense                        & Chest belt                       & HR, EDA, ECG, lung volume, breathing rate, ST                                                                                        & 1                                          & \cite{Hovsepian2015cStressTA}                                                                                                                                                                                         \\ \hline
HTC Vive Pro Eye                 & Headset                          & Eye tracking                                                                                                                                               & 1                                          & \cite{Wilson2021}                                                                                                                                                                                                     \\ \hline
BodyMedia Sensewear Pro3         & Armband                          & EDA, ECG, ST, sweat, heat flux                                                                                                               & 1                                          & \cite{Taylor2015UsingPS}                                                                                                                                                                                              \\ \hline
Zephyr Bioharness BT             & Chest strap                      & ECG, HR, breathing rate, oxygen level                                                                                                              & 1                                          & \cite{Taylor2015UsingPS}                                                                                                                                                                                              \\ \hline
Lightstone Fingertip Sensor      & Finger sensor                    & EDA, HR                                                                                                                          & 1                                          & \cite{Taylor2015UsingPS}                                                                                                                                    \\ \hline
Neulog GSR module      & Finger sensor                    & GSR                                                                                                                         & 1                                          & \cite{microstress}                                                                                                                \\ \hline
Chillband                        & Wristband                        & EDA, ST, ACC                                                                                                          & 1                                          & \cite{Yu}                                                                                                                                                                                                             \\ \hline
Healthpatch                      & Chest strap                      & ECG, ACC                                                                                                                                         & 1                                          & \cite{Yu}                                                                                                                                                                                                             \\ \hline
OMsignal                         & Body wear                        & ECG                                                                                                                                                        & 1                                          & \cite{Yu}                                                                                                                                                                                                             \\ \hline
PRO-Diary                 & Smartwatch                       & Actigraphy                                                                                                                                                 & 1                                          & \cite{Antar}                                                                                                                                                                                                           \\
\bottomrule
\end{tabular}%
}
\end{table}

While all papers introduced the device and collected signals, 31 papers failed to explicitly explain all of the used signals to the psychological constructs under investigation, especially for papers utilized smartphone sensing. Among the physiological signals, Electrodermal Activity (EDA) demonstrates promise in reflecting the activation of \textit{Sympathetic Nervous System} (SNS), mediating involuntary responses to emotion arousal, thereby serving as a potential measure for affective and cognitive states \cite{guideeda}.



\subsubsection{Data Collection}

The quality of research data is closely tied to the method of survey administration chosen. Different modes of survey administration, such as phone, online, or in-person, can introduce mode effects, where the method itself influences the responses obtained. It is crucial for researchers to report the mode of survey administration to ensure transparency and accurately interpret the findings. Surprisingly, 14 Ubicomp papers failed to report the methods of survey administration.In addition to the mode of administration, the frequency of surveys can also impact the quality of responses, particularly in natural settings. Most Ubicomp papers have wisely adopted a daily data collection approach. However, one Ubciomp paper asked participants to report 15 times a day, and another paper required participants to report five times a day. 

To mitigate bias and enhance response rates, it is crucial to provide clear guidance to participants regarding the research objectives and expectations. Only 10 Ubicomp papers reported providing such guidance to participants. Furthermore, only 3 papers reported the response rates, which is an important metric for assessing the representativeness and reliability of the collected data. 


\subsubsection{Post-processing}

Different from traditional survey research, HCC studies require special considerations of ground truth. 6 Ubicomp papers have implemented specific methods to validate the collected ground truth. For example, one paper normalized reported mood using z-score, as\textit{`some people may be more positive or negative about their mood'}. Another paper used linear interpolation computed for missing values calculation, which served as the ground truth. To ensure data quality,  11 Ubicomp papers adopted various measures. One paper \textit{`allowed participants to choose the data to collect and use'}, another one mentioned their ability to reach participants as needed. Additionally, one paper reported that \textit{`participants were unaware of the true study purpose to prevent bias'}. Another paper set a trap question with a known answer (e.g., location) and assessed reliability based on answer completion time and the trap question response. Instead, one paper accommodated participants who preferred a pen-and-paper survey over using an app, emphasizing flexibility in data collection methods. However,  most Ubicomp papers didn't transparently disclose whether specific measures were taken to ensure the quality of collected data. 
Regarding bias discussions, 22 Ubicomp papers acknowledged potential biases, such as \textit{`when participants were continuously asked to take survey they may be distracted'}, \textit{`be not sure whether the participants reported amotion truthfully or not'}, \textit{`ground truth collection may be labour intensive, expensive and somewhat inaccurate'}, etc.

\section{Reliability of Self-report as Ground Truth for Human-centred Computing}
\label{sec:reliabilty}

Recent HCC studies have highlighted significant concerns regarding the reliability of self-reported data, primarily due to improper self-reporting practices and inherent limitations in survey research methods. While it is difficult to obtain the absolute truth about human mental wellbeing, researchers have employed various methods to address these issues. These methods include investigating the misalignment between self-report and well-established technologies \cite{gao2021investigating}, comparing self-report with physiological measurements \cite{gao2021investigating}, comparing the performance of prediction models  \cite{das2022semantic}, or directly comparing self-report with real ground truths, such as monitoring security behaviours through recorded videos cite{wash2017can}.

Gao et al. \cite{gao2021investigating} conducted a study to examine the reliability of self-report measures in establishing ground truth for predicting student engagement. They discovered that the physiological measurement of engagement and the perceived engagement reported by individuals were not always consistent. This finding suggests that relying solely on subjective annotations may introduce potential unreliability when establishing ground truth. In a related study, Kaur \cite{kaur2022didn} explored the relationship between \textit{Automated Emotion Recongition} (AER) technologies and self-reported affect in the context of information work at technology companies. They revealed a misalignment between the continuous observed emotion from the AER tool and the discrete reported affect by individuals.

Das et al. \cite{semanticgap} found that when what people feel differs from what people say they feel. They identified a semantic gap that hinders accurate predictions in emotion recognition. Furthermore, they highlighted that predicting mental wellbeing using passive data, such as offline sensor data or online social media, is influenced by how the ground truth is measured, whether through objective arousal measurement or self-report. In another study, Wash et al. \cite{wash2017can} collected behavioural data and survey responses from 122 participants. They discovered that only a small number of behaviours, particularly those related to tasks that require individuals to perform specific, regular actions, exhibited non-zero correlations. Interestingly, several important security behaviours that were directly monitored did not align with self-reported responses accurately. They concluded that self-report measures are reliable only for certain behaviours. However, it is worth noting that monitoring security behaviours leans more towards physical measurements rather than psychological ones, making them easier to quantify.


\section{Improving Ground Truth Measurement in HCC Studies}
\label{sec:future directions}

Improving the practice of obtaining ground truth data is crucial for HCC studies. To achieve this, efforts have been categorized into two directions: firstly, enhancing engagement and response rate in self-report mechanism, and secondly, exploring more flexible and innovative methods for collecting ground truth data.

\subsection{Enhancing Engagement and Response Rate}

Self-report surveys, known for their convenience and cost-effectiveness, can suffer from low response rates, undermining their validity. To enhance response rates, various efforts has been made to improve engagement, such as the use of \textit{Casual Affective Triggers} (CATs) \cite{CAT}, conversational agents \cite{chatbot}, gamification \cite{van2017gamification} and incentivizing participation \cite{incspr, sprinc, incentivechi}.

Chounta et al. \cite{CAT} suggest the use of CATs, such as engaging images or interactive artifacts, in the survey interface or email notifications. They found that CATs can improve emotional engagement, thereby motivating respondents to participate and complete surveys. This strategy can lead to higher response rates and more robust, representative data.
The development of conversational agents helps to overcome low engagement by simulating human-like conversations and interacting with survey participants interactively, thereby improving the response quality \cite{chatbot}. The friendly and active nature of a conversational agent can help to create a more comfortable and personalized survey experience. Participants may feel more at ease when answering questions, leading to increased response quality and potentially reducing response bias.

Employing gamification techniques in surveys, as described by Van den Broeck et al. \cite{van2017gamification}, can significantly enhance user engagement. By incorporating game-like elements such as animations and leaderboards, the survey experience becomes more enjoyable, encouraging more accurate and reliable responses. The use of incentives to increase response and participation rates is a common practice. Studies \cite{incspr, sprinc, incentivechi} have shown that incentives and follow-up messages can improve participation rates. The Bayesian Truth Serum (BTS) technique \cite{elsincentive} addresses potential biases introduced by incentives, ensuring the reliability of responses.


\subsection{Enabling Flexible Ways for Collecting Ground Truth}
\label{subsec:flexible}
Diverse approaches exist for collecting ground truth data in the domain of mental wellbeing recognition, particularly in the areas of emotion \cite{Rodrigues, Sharma*_Mansotra_2019} and stress \cite{Bali2015ClinicalES,onlinestress}. Advances in physical and behavioural sciences have enabled the development of algorithms capable of accurately recognizing simple emotions such as happy, sad, angry, etc. These algorithms may utilize various modalities such as video \cite{zeng2007survey}, audio \cite{panda2020audio}, and text  \cite{medhat2014sentiment}, the latter often referred to as \textit{`sentiment analysis'}. Initially, these algorithms relied on annotated data as ground truth. However, due to their high accuracy in emotion detection, they have increasingly been adopted as a new form of ground truth for continuous emotion annotation in this field \cite{tag2022emotion,kaur2022didn}. 

It is crucial to acknowledge that while methods developed for emotion recognition, particularly those utilizing video and audio data, have shown promise, their effectiveness in addressing more complex psychological constructs like depression and personality traits remains limited. These algorithms identifying these more nuanced psychological states are not yet sufficient to consider them as reliable ground truth sources \cite{luxton2011mhealth, coan2007handbook}. 
This limitation indicates the need for ongoing research and development in the field. However, it also presents unique opportunities: these methods can serve as supplementary ground truth sources, contributing to a more holistic understanding of mental states and wellbeing. To provide a clearer perspective, we present a summary of the current practices in collecting ground truth data across various methods below.

\textbf{Induction Test}. Induction tests are designed to elicit physiological responses to emotions like stress, serving as a widely-accepted benchmark for establishing the ground truth about these emotional states \cite{Castaldo2017ToWE,Mozos2017StressDU,Ustress}. Key tests include the \textit{Cold Pressor Test}, \textit{Trier Social Stress Test}, \textit{Montreal Imaging Stress Task}, \textit{Maastricht Acute Stress Test}, \textit{Paced Auditory Serial Addition Task}, and \textit{Mannheim Multicomponent Stress Test} \cite{Bali2015ClinicalES}. However, to capture the complexities of real-world emotions, Almazrouei et al. \cite{onlinestress} conducted online stress induction tests for natural assessment. Additionally, Larradet et al. \cite{Larradet} introduced a mobile application linked to wearable devices for continuous monitoring of emotional states, using the \textit{Ortony, Clore, and Collins} (OCC) model to prompt users to record emotional triggers, thereby bridging the gap between lab and real-world settings.

\textbf{Vision-Based Detection}. This method utilizes facial expressions, eye movements, and physical behaviours like head movement and pupil size variation to analyze human mental states such as stress and emotions \cite{Giannakakis2017StressAA,kaur2022didn}. Some established algorithms include the \textit{Facial Action Coding System} (FACS) for categorizing facial expressions \cite{ekman1978facial}, Facebook's \textit{DeepFace} for facial recognition \cite{taigman2014deepface}, Google's \textit{Cloud Vision API} for emotional analysis in images. Due to the high accuracy of vision-based emotion detection, it is increasingly used as ground truth in HCC studies. For instance, Tag et al. \cite{tag2022emotion} utilizes the Affectiva API and the AWARE framework to monitor emotions through smartphone cameras. This application considers the captured emotions as ground truth for analyzing emotion trajectories for smartphone users. 

\textbf{Text-Based Detection}. 
Text-based emotion detection, commonly known as sentiment analysis, utilizes algorithms and natural language processing (NLP) techniques to identify emotions from textual content. The Google Cloud Natural Language API is a prominent tool in this domain, utilising machine learning for nuanced sentiment analysis. Advanced models such as BERT \cite{devlin2018bert}  have significantly improved the accuracy of emotion detection in text, making them invaluable in HCC studies as supplementary ground truth sources. For instance, Terzimehic et al. \cite{terzimehic2023tale} utilized \textit{Ortony, Clore, and Collins} (OCC) model to derive emotions from love or breakup letters to smartphones, which served as the ground truth for analyzing the emotional shifts experienced by smartphones before and during the COVID-19 pandemic.

\textbf{Speech-Based Detection}. Speech-based emotion detection differs from text-based analysis as it involves interpreting vocal cues and prosody to discern emotional states. This method analyzes the tone, pitch, and rhythm of speech, which convey a wealth of emotional information beyond the spoken words. Popular algorithms and tools such as \textit{Mel-Frequency Cepstral Coefficients} (MFCCs), OpenSMILE toolkit \cite{eyben2010opensmile} and the Emo-DB database \cite{burkhardt2005database}, have been utilized to capture the emotion during speech. 

\textbf{Multimodal Detection}. 
Multimodal approaches integrate various data types to provide a more comprehensive analysis of emotions. Sharma et al. \cite{Sharma*_Mansotra_2019} exemplify this by combining video, social media, and Twitter feedback to assess student emotions in educational settings. Non-verbal behaviours in job interviews are analyzed by studying facial expressions, speech patterns, and prosody to evaluate performance \cite{Naim}. Rodrigues et al. \cite{Rodrigues} developed a multimodal system for detecting stress and fatigue in drivers, integrating physiological, psychological, and georeferenced data. The ground truth derived from these multimodal sources enables the study of emotions in a continuous, dynamic context, offering a richer understanding of emotional states across various scenarios.

\section{Research Gaps and Future Work}
\label{sec:future work}

Through our review, we have identified the challenges in the practice of self-reporting in HCC studies. To contribute to this field, future researchers can explore several directions.

\textbf{Improve Self-Report Data Collection:} Self-report practices in HCC studies demonstrate distinct patterns compared to traditional survey research. Ubicomp researchers could consider improving the self-report data collection process by addressing factors that influence the quality and quantity of data. This can involve engaging interactions (e.g., gamification \cite{van2017gamification}, conversational chatbot \cite{chatbot}), designing better incentive mechanisms, determining optimal methods (e.g., time, frequency, location, user cognition load) for questionnaire delivery, and developing more suitable tools for HCC research, like \textit{Photographic Affect Meter} \cite{pollak2011pam}.

\textbf{Enhance Self-Report Data Quality:} The common practice in HCC studies is to use self-report data directly as ground truth without assessing its credibility. Future researchers can analyse and evaluate the quality of self-report data during the post-processing periods. This could be achieved by incorporating special and trap questions during data collection \cite{liu2018trap}, analysing completion time to assess response certainty \cite{malhotra2008completion}, measuring response biases and mitigating low-quality data, thereby enhancing overall data quality used as ground truth.


\textbf{Flexible Ground Truth Measurement:}  It is important to develop more adaptable methods for collecting ground truth data. These approaches should emphasize privacy protection, minimize the burden on users, and avoid the need for extensive installations or specialized equipment. While fully replacing self-report data may not be feasible in the short term, integrating multiple data collection modalities as discussed in Section \ref{subsec:flexible} could address some of the limitations inherent in self-reporting.

\textbf{Foster Replicable Practices:} Reproducibility is a significant challenge in HCC studies due to the nature of human-based data collection. To enhance reproducibility, it is crucial to establish standardised protocols for data collection and analysis. When collecting ground truth data, researchers should report various factors, including questionnaire delivery methods, frequency, user incentives,  guidance, etc. By focusing on these elements, researchers can increase the reproducibility and reliability of their research findings, fostering a more robust and trustworthy body of knowledge in the field.




\section{Conclusion}
\label{sec:conclusion}

This paper presents a comprehensive methodological review of Ubicomp papers that utilize self-report as the ground truth for human mental and wellbeing computing. Our analysis identifies deficiencies and inconsistencies in current self-reporting practices within the Ubicomp community. We have developed a set of guidelines that aim to improve and standardize self-reporting practices, thereby enhancing the reliability and credibility of future studies in human mental and wellbeing computing. 

Furthermore, we address the urgent need for methodological evolution and advocate for a shift from traditional self-reporting methods towards incorporating more reliable and diverse approaches, such as physiological data analysis and advanced data processing techniques. This shift is critical for advancing the accuracy and applicability of HCC research, particularly in real-world scenarios. By embracing these changes, the field of HCC can make more substantial and impactful contributions to the understanding and enhancement of human mental wellbeing.

\begin{acks}

We sincerely thank anonymous reviewers for their effort to improve the paper.
\end{acks}

\bibliographystyle{ACM-Reference-Format}
\bibliography{sample-base}


\begin{thebibliography}{160}


\ifx \showCODEN    \undefined \def \showCODEN     #1{\unskip}     \fi
\ifx \showDOI      \undefined \def \showDOI       #1{#1}\fi
\ifx \showISBNx    \undefined \def \showISBNx     #1{\unskip}     \fi
\ifx \showISBNxiii \undefined \def \showISBNxiii  #1{\unskip}     \fi
\ifx \showISSN     \undefined \def \showISSN      #1{\unskip}     \fi
\ifx \showLCCN     \undefined \def \showLCCN      #1{\unskip}     \fi
\ifx \shownote     \undefined \def \shownote      #1{#1}          \fi
\ifx \showarticletitle \undefined \def \showarticletitle #1{#1}   \fi
\ifx \showURL      \undefined \def \showURL       {\relax}        \fi
\providecommand\bibfield[2]{#2}
\providecommand\bibinfo[2]{#2}
\providecommand\natexlab[1]{#1}
\providecommand\showeprint[2][]{arXiv:#2}

\bibitem[BRO(2009)]%
        {BROCKMYER2009624}
 \bibinfo{year}{2009}\natexlab{}.
\newblock \showarticletitle{The development of the Game Engagement Questionnaire: A measure of engagement in video game-playing}.
\newblock \bibinfo{journal}{\emph{Journal of Experimental Social Psychology}} \bibinfo{volume}{45}, \bibinfo{number}{4} (\bibinfo{year}{2009}), \bibinfo{pages}{624--634}.
\newblock
\showISSN{0022-1031}
\urldef\tempurl%
\url{https://doi.org/10.1016/j.jesp.2009.02.016}
\showDOI{\tempurl}


\bibitem[Abras et~al\mbox{.}(2004)]%
        {abras2004user}
\bibfield{author}{\bibinfo{person}{Chadia Abras}, \bibinfo{person}{Diane Maloney-Krichmar}, \bibinfo{person}{Jenny Preece}, {et~al\mbox{.}}} \bibinfo{year}{2004}\natexlab{}.
\newblock \showarticletitle{User-centered design}.
\newblock \bibinfo{journal}{\emph{Bainbridge, W. Encyclopedia of Human-Computer Interaction. Thousand Oaks: Sage Publications}} \bibinfo{volume}{37}, \bibinfo{number}{4} (\bibinfo{year}{2004}), \bibinfo{pages}{445--456}.
\newblock


\bibitem[Adler et~al\mbox{.}(2021)]%
        {Identifying}
\bibfield{author}{\bibinfo{person}{Dan Adler}, \bibinfo{person}{Vincent Tseng}, \bibinfo{person}{Gengmo Qi}, \bibinfo{person}{Joseph Scarpa}, \bibinfo{person}{Srijan Sen}, {and} \bibinfo{person}{Tanzeem Choudhury}.} \bibinfo{year}{2021}\natexlab{}.
\newblock \showarticletitle{Identifying Mobile Sensing Indicators of Stress-Resilience}.
\newblock \bibinfo{journal}{\emph{Proceedings of the ACM on Interactive, Mobile, Wearable and Ubiquitous Technologies}}  \bibinfo{volume}{5} (\bibinfo{date}{06} \bibinfo{year}{2021}), \bibinfo{pages}{1--32}.
\newblock
\urldef\tempurl%
\url{https://doi.org/10.1145/3463528}
\showDOI{\tempurl}


\bibitem[Almazrouei et~al\mbox{.}(2022)]%
        {onlinestress}
\bibfield{author}{\bibinfo{person}{Mohammed Almazrouei}, \bibinfo{person}{Ruth Morgan}, {and} \bibinfo{person}{Itiel Dror}.} \bibinfo{year}{2022}\natexlab{}.
\newblock \showarticletitle{A method to induce stress in human subjects in online research environments}.
\newblock \bibinfo{journal}{\emph{Behavior Research Methods}} (\bibinfo{date}{07} \bibinfo{year}{2022}).
\newblock
\urldef\tempurl%
\url{https://doi.org/10.3758/s13428-022-01915-3}
\showDOI{\tempurl}


\bibitem[Andrews and Herzog(1986)]%
        {andrews1986quality}
\bibfield{author}{\bibinfo{person}{Frank~M Andrews} {and} \bibinfo{person}{A~Regula Herzog}.} \bibinfo{year}{1986}\natexlab{}.
\newblock \showarticletitle{The quality of survey data as related to age of respondent}.
\newblock \bibinfo{journal}{\emph{J. Amer. Statist. Assoc.}} \bibinfo{volume}{81}, \bibinfo{number}{394} (\bibinfo{year}{1986}), \bibinfo{pages}{403--410}.
\newblock


\bibitem[Antar et~al\mbox{.}(2023)]%
        {Antar}
\bibfield{author}{\bibinfo{person}{Anindya~Das Antar}, \bibinfo{person}{Anna Kratz}, {and} \bibinfo{person}{Nikola Banovic}.} \bibinfo{year}{2023}\natexlab{}.
\newblock \showarticletitle{Behavior Modeling Approach for Forecasting Physical Functioning of People with Multiple Sclerosis}.
\newblock \bibinfo{journal}{\emph{Proc. ACM Interact. Mob. Wearable Ubiquitous Technol.}} \bibinfo{volume}{7}, \bibinfo{number}{1}, Article \bibinfo{articleno}{7} (\bibinfo{date}{mar} \bibinfo{year}{2023}), \bibinfo{numpages}{29}~pages.
\newblock
\urldef\tempurl%
\url{https://doi.org/10.1145/3580887}
\showDOI{\tempurl}


\bibitem[Arakawa et~al\mbox{.}(2023)]%
        {Arkawa}
\bibfield{author}{\bibinfo{person}{Riku Arakawa}, \bibinfo{person}{Karan Ahuja}, \bibinfo{person}{Kristie Mak}, \bibinfo{person}{Gwendolyn Thompson}, \bibinfo{person}{Sam Shaaban}, \bibinfo{person}{Oliver Lindhiem}, {and} \bibinfo{person}{Mayank Goel}.} \bibinfo{year}{2023}\natexlab{}.
\newblock \showarticletitle{LemurDx: Using Unconstrained Passive Sensing for an Objective Measurement of Hyperactivity in Children with No Parent Input}.
\newblock \bibinfo{journal}{\emph{Proc. ACM Interact. Mob. Wearable Ubiquitous Technol.}} \bibinfo{volume}{7}, \bibinfo{number}{2}, Article \bibinfo{articleno}{46} (\bibinfo{date}{jun} \bibinfo{year}{2023}), \bibinfo{numpages}{23}~pages.
\newblock
\urldef\tempurl%
\url{https://doi.org/10.1145/3596244}
\showDOI{\tempurl}


\bibitem[Attal et~al\mbox{.}(2015)]%
        {attal2015physical}
\bibfield{author}{\bibinfo{person}{Ferhat Attal}, \bibinfo{person}{Samer Mohammed}, \bibinfo{person}{Mariam Dedabrishvili}, \bibinfo{person}{Faicel Chamroukhi}, \bibinfo{person}{Latifa Oukhellou}, {and} \bibinfo{person}{Yacine Amirat}.} \bibinfo{year}{2015}\natexlab{}.
\newblock \showarticletitle{Physical human activity recognition using wearable sensors}.
\newblock \bibinfo{journal}{\emph{Sensors}} \bibinfo{volume}{15}, \bibinfo{number}{12} (\bibinfo{year}{2015}), \bibinfo{pages}{31314--31338}.
\newblock


\bibitem[Baillon et~al\mbox{.}(2022)]%
        {elsincentive}
\bibfield{author}{\bibinfo{person}{Aurélien Baillon}, \bibinfo{person}{Han Bleichrodt}, {and} \bibinfo{person}{Georg~D. Granic}.} \bibinfo{year}{2022}\natexlab{}.
\newblock \showarticletitle{Incentives in surveys}.
\newblock \bibinfo{journal}{\emph{Journal of Economic Psychology}}  \bibinfo{volume}{93} (\bibinfo{year}{2022}), \bibinfo{pages}{102552}.
\newblock
\showISSN{0167-4870}
\urldef\tempurl%
\url{https://doi.org/10.1016/j.joep.2022.102552}
\showDOI{\tempurl}


\bibitem[Bali and Jaggi(2015)]%
        {Bali2015ClinicalES}
\bibfield{author}{\bibinfo{person}{Anjana Bali} {and} \bibinfo{person}{Amteshwar~Singh Jaggi}.} \bibinfo{year}{2015}\natexlab{}.
\newblock \showarticletitle{Clinical experimental stress studies: methods and assessment}.
\newblock \bibinfo{journal}{\emph{Reviews in the Neurosciences}}  \bibinfo{volume}{26} (\bibinfo{year}{2015}), \bibinfo{pages}{555 -- 579}.
\newblock


\bibitem[Baumeister and Leary(1995)]%
        {loneliness}
\bibfield{author}{\bibinfo{person}{Roy Baumeister} {and} \bibinfo{person}{Mark Leary}.} \bibinfo{year}{1995}\natexlab{}.
\newblock \showarticletitle{The Need to Belong: Desire for Interpersonal Attachments as a Fundamental Human Motivation}.
\newblock \bibinfo{journal}{\emph{Psychological bulletin}}  \bibinfo{volume}{117} (\bibinfo{date}{06} \bibinfo{year}{1995}), \bibinfo{pages}{497--529}.
\newblock
\urldef\tempurl%
\url{https://doi.org/10.1037/0033-2909.117.3.497}
\showDOI{\tempurl}


\bibitem[Beck et~al\mbox{.}(1996)]%
        {Beck1996ComparisonOB}
\bibfield{author}{\bibinfo{person}{Aaron~T. Beck}, \bibinfo{person}{Robert~A. Steer}, \bibinfo{person}{Roberta Ball}, {and} \bibinfo{person}{William~F. Ranieri}.} \bibinfo{year}{1996}\natexlab{}.
\newblock \showarticletitle{Comparison of Beck Depression Inventories -IA and -II in psychiatric outpatients.}
\newblock \bibinfo{journal}{\emph{Journal of personality assessment}}  \bibinfo{volume}{67 3} (\bibinfo{year}{1996}), \bibinfo{pages}{588--97}.
\newblock


\bibitem[Biel et~al\mbox{.}(2012)]%
        {biel2012facetube}
\bibfield{author}{\bibinfo{person}{Joan-Isaac Biel}, \bibinfo{person}{Luc{\'\i}a Teijeiro-Mosquera}, {and} \bibinfo{person}{Daniel Gatica-Perez}.} \bibinfo{year}{2012}\natexlab{}.
\newblock \showarticletitle{Facetube: predicting personality from facial expressions of emotion in online conversational video}. In \bibinfo{booktitle}{\emph{Proceedings of the 14th ACM international conference on Multimodal interaction}}. \bibinfo{pages}{53--56}.
\newblock


\bibitem[Bos et~al\mbox{.}(2015)]%
        {bos2015experience}
\bibfield{author}{\bibinfo{person}{Fionneke~M Bos}, \bibinfo{person}{Robert~A Schoevers}, {and} \bibinfo{person}{Marije aan~het Rot}.} \bibinfo{year}{2015}\natexlab{}.
\newblock \showarticletitle{Experience sampling and ecological momentary assessment studies in psychopharmacology: a systematic review}.
\newblock \bibinfo{journal}{\emph{European Neuropsychopharmacology}} \bibinfo{volume}{25}, \bibinfo{number}{11} (\bibinfo{year}{2015}), \bibinfo{pages}{1853--1864}.
\newblock


\bibitem[Braithwaite et~al\mbox{.}(2013)]%
        {guideeda}
\bibfield{author}{\bibinfo{person}{Jason~J Braithwaite}, \bibinfo{person}{Derrick~G Watson}, \bibinfo{person}{Robert Jones}, {and} \bibinfo{person}{Mickey Rowe}.} \bibinfo{year}{2013}\natexlab{}.
\newblock \showarticletitle{A guide for analysing electrodermal activity (EDA) \& skin conductance responses (SCRs) for psychological experiments}.
\newblock \bibinfo{journal}{\emph{Psychophysiology}} \bibinfo{volume}{49}, \bibinfo{number}{1} (\bibinfo{year}{2013}), \bibinfo{pages}{1017--1034}.
\newblock


\bibitem[Burkhardt et~al\mbox{.}(2005)]%
        {burkhardt2005database}
\bibfield{author}{\bibinfo{person}{Felix Burkhardt}, \bibinfo{person}{Astrid Paeschke}, \bibinfo{person}{Miriam Rolfes}, \bibinfo{person}{Walter~F Sendlmeier}, \bibinfo{person}{Benjamin Weiss}, {et~al\mbox{.}}} \bibinfo{year}{2005}\natexlab{}.
\newblock \showarticletitle{A database of German emotional speech.}. In \bibinfo{booktitle}{\emph{Interspeech}}, Vol.~\bibinfo{volume}{5}. \bibinfo{pages}{1517--1520}.
\newblock


\bibitem[Canzian and Musolesi(2015)]%
        {Canzian2015TrajectoriesOD}
\bibfield{author}{\bibinfo{person}{Luca Canzian} {and} \bibinfo{person}{Mirco Musolesi}.} \bibinfo{year}{2015}\natexlab{}.
\newblock \showarticletitle{Trajectories of depression: unobtrusive monitoring of depressive states by means of smartphone mobility traces analysis}.
\newblock \bibinfo{journal}{\emph{Proceedings of the 2015 ACM International Joint Conference on Pervasive and Ubiquitous Computing}} (\bibinfo{year}{2015}).
\newblock


\bibitem[Carlozzi(2011)]%
        {Carlozzi2011}
\bibfield{author}{\bibinfo{person}{Noelle~E. Carlozzi}.} \bibinfo{year}{2011}\natexlab{}.
\newblock \bibinfo{booktitle}{\emph{Shipley Institute of Living Scale}}.
\newblock \bibinfo{publisher}{Springer New York}, \bibinfo{address}{New York, NY}, \bibinfo{pages}{2287--2289}.
\newblock
\showISBNx{978-0-387-79948-3}
\urldef\tempurl%
\url{https://doi.org/10.1007/978-0-387-79948-3_1070}
\showDOI{\tempurl}


\bibitem[Castaldo et~al\mbox{.}(2017)]%
        {Castaldo2017ToWE}
\bibfield{author}{\bibinfo{person}{Rossana Castaldo}, \bibinfo{person}{Luis Montesinos}, \bibinfo{person}{Paolo Melillo}, \bibinfo{person}{Sebastiano Massaro}, {and} \bibinfo{person}{Leandro Pecchia}.} \bibinfo{year}{2017}\natexlab{}.
\newblock \showarticletitle{To What Extent Can We Shorten HRV Analysis in Wearable Sensing? A Case Study on Mental Stress Detection.}
\newblock


\bibitem[Cella et~al\mbox{.}(2012)]%
        {Cella1860}
\bibfield{author}{\bibinfo{person}{D. Cella}, \bibinfo{person}{J.-S. Lai}, \bibinfo{person}{C.J. Nowinski}, \bibinfo{person}{D. Victorson}, \bibinfo{person}{A. Peterman}, \bibinfo{person}{D. Miller}, \bibinfo{person}{F. Bethoux}, \bibinfo{person}{A. Heinemann}, \bibinfo{person}{S. Rubin}, \bibinfo{person}{J.E. Cavazos}, \bibinfo{person}{A.T. Reder}, \bibinfo{person}{R. Sufit}, \bibinfo{person}{T. Simuni}, \bibinfo{person}{G.L. Holmes}, \bibinfo{person}{A. Siderowf}, \bibinfo{person}{V. Wojna}, \bibinfo{person}{R. Bode}, \bibinfo{person}{N. McKinney}, \bibinfo{person}{T. Podrabsky}, \bibinfo{person}{K. Wortman}, \bibinfo{person}{S. Choi}, \bibinfo{person}{R. Gershon}, \bibinfo{person}{N. Rothrock}, {and} \bibinfo{person}{C. Moy}.} \bibinfo{year}{2012}\natexlab{}.
\newblock \showarticletitle{Neuro-QOL}.
\newblock \bibinfo{journal}{\emph{Neurology}} \bibinfo{volume}{78}, \bibinfo{number}{23} (\bibinfo{year}{2012}), \bibinfo{pages}{1860--1867}.
\newblock
\showISSN{0028-3878}
\urldef\tempurl%
\url{https://doi.org/10.1212/WNL.0b013e318258f744}
\showDOI{\tempurl}
\showeprint{https://n.neurology.org/content/78/23/1860.full.pdf}


\bibitem[Choi and Pak(2004)]%
        {Choi2004ACO}
\bibfield{author}{\bibinfo{person}{Bernard C.~K. Choi} {and} \bibinfo{person}{Anita W.~P. Pak}.} \bibinfo{year}{2004}\natexlab{}.
\newblock \showarticletitle{A Catalog of Biases in Questionnaires}.
\newblock \bibinfo{journal}{\emph{Preventing Chronic Disease}}  \bibinfo{volume}{2} (\bibinfo{year}{2004}).
\newblock


\bibitem[Chounta and Nolte(2022)]%
        {CAT}
\bibfield{author}{\bibinfo{person}{Irene-Angelica Chounta} {and} \bibinfo{person}{Alexander Nolte}.} \bibinfo{year}{2022}\natexlab{}.
\newblock \showarticletitle{The CAT Effect: Exploring the Impact of Casual Affective Triggers on Online Surveys’ Response Rates} \emph{(\bibinfo{series}{CHI '22})}. \bibinfo{publisher}{Association for Computing Machinery}, \bibinfo{address}{New York, NY, USA}, Article \bibinfo{articleno}{583}, \bibinfo{numpages}{13}~pages.
\newblock
\showISBNx{9781450391573}
\urldef\tempurl%
\url{https://doi.org/10.1145/3491102.3517481}
\showDOI{\tempurl}


\bibitem[Coan and Allen(2007)]%
        {coan2007handbook}
\bibfield{author}{\bibinfo{person}{James~A Coan} {and} \bibinfo{person}{John~JB Allen}.} \bibinfo{year}{2007}\natexlab{}.
\newblock \bibinfo{booktitle}{\emph{Handbook of emotion elicitation and assessment}}.
\newblock \bibinfo{publisher}{Oxford university press}.
\newblock


\bibitem[Cohen et~al\mbox{.}(1983)]%
        {Cohen1983AGM}
\bibfield{author}{\bibinfo{person}{Sheldon Cohen}, \bibinfo{person}{Tom~P Kamarck}, {and} \bibinfo{person}{Robin~J. Mermelstein}.} \bibinfo{year}{1983}\natexlab{}.
\newblock \showarticletitle{A global measure of perceived stress.}
\newblock \bibinfo{journal}{\emph{Journal of health and social behavior}}  \bibinfo{volume}{24 4} (\bibinfo{year}{1983}), \bibinfo{pages}{385--96}.
\newblock


\bibitem[Cooper(2001)]%
        {cooper2001diagnostic}
\bibfield{author}{\bibinfo{person}{John Cooper}.} \bibinfo{year}{2001}\natexlab{}.
\newblock \showarticletitle{Diagnostic and statistical manual of mental disorders (4th edn, text revision)(DSM--IV--TR) Washington, DC: American Psychiatric Association 2000. 943 pp.{\pounds} 39.99 (hb). ISBN 0 89042 025 4}.
\newblock \bibinfo{journal}{\emph{The British Journal of Psychiatry}} \bibinfo{volume}{179}, \bibinfo{number}{1} (\bibinfo{year}{2001}), \bibinfo{pages}{85--85}.
\newblock


\bibitem[Costa et~al\mbox{.}(2016)]%
        {Costa2016EmotionCheckLB}
\bibfield{author}{\bibinfo{person}{Jean Dos~Reis Costa}, \bibinfo{person}{Alexander~Travis Adams}, \bibinfo{person}{Malte~F. Jung}, \bibinfo{person}{François Guimbreti{\`e}re}, {and} \bibinfo{person}{Tanzeem Choudhury}.} \bibinfo{year}{2016}\natexlab{}.
\newblock \showarticletitle{EmotionCheck: leveraging bodily signals and false feedback to regulate our emotions}.
\newblock \bibinfo{journal}{\emph{Proceedings of the 2016 ACM International Joint Conference on Pervasive and Ubiquitous Computing}} (\bibinfo{year}{2016}).
\newblock


\bibitem[Crawford and Henry(2004)]%
        {Crawford2004ThePA}
\bibfield{author}{\bibinfo{person}{John~R. Crawford} {and} \bibinfo{person}{Julie~D. Henry}.} \bibinfo{year}{2004}\natexlab{}.
\newblock \showarticletitle{The positive and negative affect schedule (PANAS): construct validity, measurement properties and normative data in a large non-clinical sample.}
\newblock \bibinfo{journal}{\emph{The British journal of clinical psychology}}  \bibinfo{volume}{43 Pt 3} (\bibinfo{year}{2004}), \bibinfo{pages}{245--65}.
\newblock


\bibitem[Das~Swain et~al\mbox{.}(2022a)]%
        {semanticgap}
\bibfield{author}{\bibinfo{person}{Vedant Das~Swain}, \bibinfo{person}{Victor Chen}, \bibinfo{person}{Shrija Mishra}, \bibinfo{person}{Stephen~M. Mattingly}, \bibinfo{person}{Gregory~D. Abowd}, {and} \bibinfo{person}{Munmun De~Choudhury}.} \bibinfo{year}{2022}\natexlab{a}.
\newblock \showarticletitle{Semantic Gap in Predicting Mental Wellbeing through Passive Sensing}. In \bibinfo{booktitle}{\emph{Proceedings of the 2022 CHI Conference on Human Factors in Computing Systems}} (New Orleans, LA, USA) \emph{(\bibinfo{series}{CHI '22})}. \bibinfo{publisher}{Association for Computing Machinery}, \bibinfo{address}{New York, NY, USA}, Article \bibinfo{articleno}{374}, \bibinfo{numpages}{16}~pages.
\newblock
\showISBNx{9781450391573}
\urldef\tempurl%
\url{https://doi.org/10.1145/3491102.3502037}
\showDOI{\tempurl}


\bibitem[Das~Swain et~al\mbox{.}(2022b)]%
        {das2022semantic}
\bibfield{author}{\bibinfo{person}{Vedant Das~Swain}, \bibinfo{person}{Victor Chen}, \bibinfo{person}{Shrija Mishra}, \bibinfo{person}{Stephen~M Mattingly}, \bibinfo{person}{Gregory~D Abowd}, {and} \bibinfo{person}{Munmun De~Choudhury}.} \bibinfo{year}{2022}\natexlab{b}.
\newblock \showarticletitle{Semantic Gap in Predicting Mental Wellbeing through Passive Sensing}. In \bibinfo{booktitle}{\emph{Proceedings of the 2022 CHI Conference on Human Factors in Computing Systems}}. \bibinfo{pages}{1--16}.
\newblock


\bibitem[Davies(2020)]%
        {bias}
\bibfield{author}{\bibinfo{person}{R.~S. Davies}.} \bibinfo{year}{2020}\natexlab{}.
\newblock \bibinfo{booktitle}{\emph{Designing Surveys for Evaluations and Research}}.
\newblock
\urldef\tempurl%
\url{https://edtechbooks.org/designing_surveys}
\showURL{%
\tempurl}


\bibitem[De~Leeuw et~al\mbox{.}(2012)]%
        {de2012international}
\bibfield{author}{\bibinfo{person}{Edith~D De~Leeuw}, \bibinfo{person}{Joop Hox}, {and} \bibinfo{person}{Don Dillman}.} \bibinfo{year}{2012}\natexlab{}.
\newblock \bibinfo{booktitle}{\emph{International handbook of survey methodology}}.
\newblock \bibinfo{publisher}{Routledge}.
\newblock


\bibitem[de~Vreede et~al\mbox{.}(2019)]%
        {engagement1}
\bibfield{author}{\bibinfo{person}{Triparna de Vreede}, \bibinfo{person}{Stephanie Andel}, \bibinfo{person}{Gert-Jan de Vreede}, \bibinfo{person}{Paul Spector}, \bibinfo{person}{Vivek Singh}, {and} \bibinfo{person}{Balaji Padmanabhan}.} \bibinfo{year}{2019}\natexlab{}.
\newblock \showarticletitle{What is Engagement and How Do We Measure It? Toward a Domain Independent Definition and Scale}.
\newblock
\urldef\tempurl%
\url{https://doi.org/10.24251/HICSS.2019.092}
\showDOI{\tempurl}


\bibitem[Depression(2012)]%
        {depression2012depression}
\bibfield{author}{\bibinfo{person}{What~Causes Depression}.} \bibinfo{year}{2012}\natexlab{}.
\newblock \showarticletitle{what is depression?}
\newblock \bibinfo{journal}{\emph{World Health Organization}} (\bibinfo{year}{2012}).
\newblock


\bibitem[Devlin et~al\mbox{.}(2018)]%
        {devlin2018bert}
\bibfield{author}{\bibinfo{person}{Jacob Devlin}, \bibinfo{person}{Ming-Wei Chang}, \bibinfo{person}{Kenton Lee}, {and} \bibinfo{person}{Kristina Toutanova}.} \bibinfo{year}{2018}\natexlab{}.
\newblock \showarticletitle{Bert: Pre-training of deep bidirectional transformers for language understanding}.
\newblock \bibinfo{journal}{\emph{arXiv preprint arXiv:1810.04805}} (\bibinfo{year}{2018}).
\newblock


\bibitem[Devos et~al\mbox{.}(2020)]%
        {article3}
\bibfield{author}{\bibinfo{person}{Hannes Devos}, \bibinfo{person}{Kathleen Gustafson}, \bibinfo{person}{Pedram Ahmadnezhad}, \bibinfo{person}{Ke Liao}, \bibinfo{person}{Jonathan Mahnken}, \bibinfo{person}{William Brooks}, {and} \bibinfo{person}{Jeffrey Burns}.} \bibinfo{year}{2020}\natexlab{}.
\newblock \showarticletitle{Psychometric Properties of NASA-TLX and Index of Cognitive Activity as Measures of Cognitive Workload in Older Adults}.
\newblock \bibinfo{journal}{\emph{Brain Sciences}}  \bibinfo{volume}{10} (\bibinfo{date}{12} \bibinfo{year}{2020}), \bibinfo{pages}{994}.
\newblock
\urldef\tempurl%
\url{https://doi.org/10.3390/brainsci10120994}
\showDOI{\tempurl}


\bibitem[Diener et~al\mbox{.}(2010)]%
        {articleflo}
\bibfield{author}{\bibinfo{person}{Ed Diener}, \bibinfo{person}{Derrick Wirtz}, {and} \bibinfo{person}{William Tov}.} \bibinfo{year}{2010}\natexlab{}.
\newblock \showarticletitle{New measures of well-being: Flourishing and positive and negative feelings}.
\newblock \bibinfo{journal}{\emph{Soc Indic Res}}  \bibinfo{volume}{39} (\bibinfo{date}{01} \bibinfo{year}{2010}), \bibinfo{pages}{247--266}.
\newblock


\bibitem[DiSalvo et~al\mbox{.}(2022a)]%
        {disalvo2022reading}
\bibfield{author}{\bibinfo{person}{Betsy DiSalvo}, \bibinfo{person}{Dheeraj Bandaru}, \bibinfo{person}{Qiaosi Wang}, \bibinfo{person}{Hong Li}, {and} \bibinfo{person}{Thomas Pl{\"o}tz}.} \bibinfo{year}{2022}\natexlab{a}.
\newblock \showarticletitle{Reading the Room: Automated, Momentary Assessment of Student Engagement in the Classroom: Are We There Yet?}
\newblock \bibinfo{journal}{\emph{Proceedings of the ACM on Interactive, Mobile, Wearable and Ubiquitous Technologies}} \bibinfo{volume}{6}, \bibinfo{number}{3} (\bibinfo{year}{2022}), \bibinfo{pages}{1--26}.
\newblock


\bibitem[DiSalvo et~al\mbox{.}(2022b)]%
        {Reading}
\bibfield{author}{\bibinfo{person}{Betsy DiSalvo}, \bibinfo{person}{Dheeraj Bandaru}, \bibinfo{person}{Qiaosi Wang}, \bibinfo{person}{Hong Li}, {and} \bibinfo{person}{Thomas Pl\"{o}tz}.} \bibinfo{year}{2022}\natexlab{b}.
\newblock \showarticletitle{Reading the Room: Automated, Momentary Assessment of Student Engagement in the Classroom: Are We There Yet?}
\newblock \bibinfo{journal}{\emph{Proc. ACM Interact. Mob. Wearable Ubiquitous Technol.}} \bibinfo{volume}{6}, \bibinfo{number}{3}, Article \bibinfo{articleno}{112} (\bibinfo{date}{sep} \bibinfo{year}{2022}), \bibinfo{numpages}{26}~pages.
\newblock
\urldef\tempurl%
\url{https://doi.org/10.1145/3550328}
\showDOI{\tempurl}


\bibitem[Dzedzickis et~al\mbox{.}(2020)]%
        {dzedzickis2020human}
\bibfield{author}{\bibinfo{person}{Andrius Dzedzickis}, \bibinfo{person}{Art{\=u}ras Kaklauskas}, {and} \bibinfo{person}{Vytautas Bucinskas}.} \bibinfo{year}{2020}\natexlab{}.
\newblock \showarticletitle{Human emotion recognition: Review of sensors and methods}.
\newblock \bibinfo{journal}{\emph{Sensors}} \bibinfo{volume}{20}, \bibinfo{number}{3} (\bibinfo{year}{2020}), \bibinfo{pages}{592}.
\newblock


\bibitem[Egilmez et~al\mbox{.}(2017)]%
        {Ustress}
\bibfield{author}{\bibinfo{person}{Begum Egilmez}, \bibinfo{person}{Emirhan Poyraz}, \bibinfo{person}{Wenting Zhou}, \bibinfo{person}{Gokhan Memik}, \bibinfo{person}{Peter Dinda}, {and} \bibinfo{person}{Nabil Alshurafa}.} \bibinfo{year}{2017}\natexlab{}.
\newblock \showarticletitle{UStress: Understanding college student subjective stress using wrist-based passive sensing}. In \bibinfo{booktitle}{\emph{2017 IEEE International Conference on Pervasive Computing and Communications Workshops (PerCom Workshops)}}. \bibinfo{pages}{673--678}.
\newblock
\urldef\tempurl%
\url{https://doi.org/10.1109/PERCOMW.2017.7917644}
\showDOI{\tempurl}


\bibitem[Ekkekakis(2013)]%
        {ekkekakis2013measurement}
\bibfield{author}{\bibinfo{person}{Panteleimon Ekkekakis}.} \bibinfo{year}{2013}\natexlab{}.
\newblock \bibinfo{booktitle}{\emph{The measurement of affect, mood, and emotion: A guide for health-behavioral research}}.
\newblock \bibinfo{publisher}{Cambridge University Press}.
\newblock


\bibitem[Ekman and Friesen(1978)]%
        {ekman1978facial}
\bibfield{author}{\bibinfo{person}{Paul Ekman} {and} \bibinfo{person}{Wallace~V Friesen}.} \bibinfo{year}{1978}\natexlab{}.
\newblock \showarticletitle{Facial action coding system}.
\newblock \bibinfo{journal}{\emph{Environmental Psychology \& Nonverbal Behavior}} (\bibinfo{year}{1978}).
\newblock


\bibitem[Exler et~al\mbox{.}(2016)]%
        {Exler2016AWS}
\bibfield{author}{\bibinfo{person}{Anja Exler}, \bibinfo{person}{Andrea Schankin}, \bibinfo{person}{Christoph Klebsattel}, {and} \bibinfo{person}{Michael Beigl}.} \bibinfo{year}{2016}\natexlab{}.
\newblock \showarticletitle{A wearable system for mood assessment considering smartphone features and data from mobile ECGs}.
\newblock \bibinfo{journal}{\emph{Proceedings of the 2016 ACM International Joint Conference on Pervasive and Ubiquitous Computing: Adjunct}} (\bibinfo{year}{2016}).
\newblock


\bibitem[Eyben et~al\mbox{.}(2010)]%
        {eyben2010opensmile}
\bibfield{author}{\bibinfo{person}{Florian Eyben}, \bibinfo{person}{Martin W{\"o}llmer}, {and} \bibinfo{person}{Bj{\"o}rn Schuller}.} \bibinfo{year}{2010}\natexlab{}.
\newblock \showarticletitle{Opensmile: the munich versatile and fast open-source audio feature extractor}. In \bibinfo{booktitle}{\emph{Proceedings of the 18th ACM international conference on Multimedia}}. \bibinfo{pages}{1459--1462}.
\newblock


\bibitem[Fink(2003)]%
        {fink2003survey}
\bibfield{author}{\bibinfo{person}{Arlene Fink}.} \bibinfo{year}{2003}\natexlab{}.
\newblock \bibinfo{booktitle}{\emph{The survey handbook}}.
\newblock \bibinfo{publisher}{sage}.
\newblock


\bibitem[Fink(2015)]%
        {fink2015conduct}
\bibfield{author}{\bibinfo{person}{Arlene Fink}.} \bibinfo{year}{2015}\natexlab{}.
\newblock \bibinfo{booktitle}{\emph{How to conduct surveys: A step-by-step guide}}.
\newblock \bibinfo{publisher}{Sage Publications}.
\newblock


\bibitem[Fiore et~al\mbox{.}(2014)]%
        {incentivechi}
\bibfield{author}{\bibinfo{person}{Andrew~T. Fiore}, \bibinfo{person}{Coye Cheshire}, \bibinfo{person}{Lindsay Shaw~Taylor}, {and} \bibinfo{person}{G.A. Mendelsohn}.} \bibinfo{year}{2014}\natexlab{}.
\newblock \showarticletitle{Incentives to Participate in Online Research: An Experimental Examination of "Surprise" Incentives}. In \bibinfo{booktitle}{\emph{Proceedings of the SIGCHI Conference on Human Factors in Computing Systems}} (Toronto, Ontario, Canada) \emph{(\bibinfo{series}{CHI '14})}. \bibinfo{publisher}{Association for Computing Machinery}, \bibinfo{address}{New York, NY, USA}, \bibinfo{pages}{3433–3442}.
\newblock
\showISBNx{9781450324731}
\urldef\tempurl%
\url{https://doi.org/10.1145/2556288.2557418}
\showDOI{\tempurl}


\bibitem[Fox et~al\mbox{.}(2000)]%
        {fox}
\bibfield{author}{\bibinfo{person}{Nick Fox}, \bibinfo{person}{Nigel Mathers}, {and} \bibinfo{person}{Amanda Hunn}.} \bibinfo{year}{2000}\natexlab{}.
\newblock \bibinfo{booktitle}{\emph{Surveys and Questionnaires}}.
\newblock \bibinfo{pages}{77--112}.
\newblock
\showISBNx{1 85775 3923}


\bibitem[French(1981)]%
        {french1981methodological}
\bibfield{author}{\bibinfo{person}{Kate French}.} \bibinfo{year}{1981}\natexlab{}.
\newblock \showarticletitle{Methodological considerations in hospital patient opinion surveys}.
\newblock \bibinfo{journal}{\emph{International journal of nursing studies}} \bibinfo{volume}{18}, \bibinfo{number}{1} (\bibinfo{year}{1981}), \bibinfo{pages}{7--32}.
\newblock


\bibitem[Fuller et~al\mbox{.}(2018)]%
        {fuller2018development}
\bibfield{author}{\bibinfo{person}{Kathryn~A Fuller}, \bibinfo{person}{Nilushi~S Karunaratne}, \bibinfo{person}{Som Naidu}, \bibinfo{person}{Betty Exintaris}, \bibinfo{person}{Jennifer~L Short}, \bibinfo{person}{Michael~D Wolcott}, \bibinfo{person}{Scott Singleton}, {and} \bibinfo{person}{Paul~J White}.} \bibinfo{year}{2018}\natexlab{}.
\newblock \showarticletitle{Development of a Self-report Instrument for Measuring in-class Student Engagement Reveals that Pretending to Engage is a Significant Unrecognized Problem}.
\newblock \bibinfo{journal}{\emph{PLoS ONE}} \bibinfo{volume}{13}, \bibinfo{number}{10} (\bibinfo{year}{2018}), \bibinfo{pages}{e0205828}.
\newblock


\bibitem[Furnham and Henderson(1982)]%
        {furnham1982good}
\bibfield{author}{\bibinfo{person}{Adrian Furnham} {and} \bibinfo{person}{Monika Henderson}.} \bibinfo{year}{1982}\natexlab{}.
\newblock \showarticletitle{The good, the bad and the mad: Response bias in self-report measures}.
\newblock \bibinfo{journal}{\emph{Personality and Individual Differences}} \bibinfo{volume}{3}, \bibinfo{number}{3} (\bibinfo{year}{1982}), \bibinfo{pages}{311--320}.
\newblock


\bibitem[Galy et~al\mbox{.}(2011)]%
        {article2}
\bibfield{author}{\bibinfo{person}{Edith Galy}, \bibinfo{person}{Magali Cariou}, {and} \bibinfo{person}{Claudine Mélan}.} \bibinfo{year}{2011}\natexlab{}.
\newblock \showarticletitle{What is the relationship between mental workload factors and cognitive load types?}
\newblock \bibinfo{journal}{\emph{International journal of psychophysiology : official journal of the International Organization of Psychophysiology}}  \bibinfo{volume}{83} (\bibinfo{date}{10} \bibinfo{year}{2011}), \bibinfo{pages}{269--75}.
\newblock
\urldef\tempurl%
\url{https://doi.org/10.1016/j.ijpsycho.2011.09.023}
\showDOI{\tempurl}


\bibitem[Gao et~al\mbox{.}(2022)]%
        {gao2022individual}
\bibfield{author}{\bibinfo{person}{Nan Gao}, \bibinfo{person}{Mohammad~Saiedur Rahaman}, \bibinfo{person}{Wei Shao}, \bibinfo{person}{Kaixin Ji}, {and} \bibinfo{person}{Flora~D Salim}.} \bibinfo{year}{2022}\natexlab{}.
\newblock \showarticletitle{Individual and group-wise classroom seating experience: Effects on student engagement in different courses}.
\newblock \bibinfo{journal}{\emph{Proceedings of the ACM on Interactive, Mobile, Wearable and Ubiquitous Technologies}} \bibinfo{volume}{6}, \bibinfo{number}{3} (\bibinfo{year}{2022}), \bibinfo{pages}{1--23}.
\newblock


\bibitem[Gao et~al\mbox{.}(2021)]%
        {gao2021investigating}
\bibfield{author}{\bibinfo{person}{Nan Gao}, \bibinfo{person}{Mohammad Saiedur~Rahaman}, \bibinfo{person}{Wei Shao}, {and} \bibinfo{person}{Flora~D Salim}.} \bibinfo{year}{2021}\natexlab{}.
\newblock \showarticletitle{Investigating the reliability of self-report data in the wild: The quest for ground truth}. In \bibinfo{booktitle}{\emph{Adjunct Proceedings of the 2021 ACM International Joint Conference on Pervasive and Ubiquitous Computing and Proceedings of the 2021 ACM International Symposium on Wearable Computers}}. \bibinfo{pages}{237--242}.
\newblock


\bibitem[Gao et~al\mbox{.}(2020)]%
        {gao2020n}
\bibfield{author}{\bibinfo{person}{Nan Gao}, \bibinfo{person}{Wei Shao}, \bibinfo{person}{Mohammad~Saiedur Rahaman}, {and} \bibinfo{person}{Flora~D Salim}.} \bibinfo{year}{2020}\natexlab{}.
\newblock \showarticletitle{n-Gage: Predicting in-class Emotional, Behavioural and Cognitive Engagement in the Wild}.
\newblock \bibinfo{journal}{\emph{Proceedings of the ACM on Interactive, Mobile, Wearable and Ubiquitous Technologies}} \bibinfo{volume}{4}, \bibinfo{number}{3} (\bibinfo{year}{2020}), \bibinfo{pages}{1--26}.
\newblock


\bibitem[Gao et~al\mbox{.}(2019)]%
        {gao2019predicting}
\bibfield{author}{\bibinfo{person}{Nan Gao}, \bibinfo{person}{Wei Shao}, {and} \bibinfo{person}{Flora~D Salim}.} \bibinfo{year}{2019}\natexlab{}.
\newblock \showarticletitle{Predicting Personality Traits from Physical Activity Intensity}.
\newblock \bibinfo{journal}{\emph{Computer}} \bibinfo{volume}{52}, \bibinfo{number}{7} (\bibinfo{year}{2019}), \bibinfo{pages}{47--56}.
\newblock


\bibitem[Gashi et~al\mbox{.}(2019)]%
        {Gashi2019UsingUW}
\bibfield{author}{\bibinfo{person}{Shkurta Gashi}, \bibinfo{person}{Elena~Di Lascio}, {and} \bibinfo{person}{Silvia Santini}.} \bibinfo{year}{2019}\natexlab{}.
\newblock \showarticletitle{Using Unobtrusive Wearable Sensors to Measure the Physiological Synchrony Between Presenters and Audience Members}.
\newblock \bibinfo{journal}{\emph{Proceedings of the ACM on Interactive, Mobile, Wearable and Ubiquitous Technologies}}  \bibinfo{volume}{3} (\bibinfo{year}{2019}), \bibinfo{pages}{1 -- 19}.
\newblock


\bibitem[Giannakakis et~al\mbox{.}(2017)]%
        {Giannakakis2017StressAA}
\bibfield{author}{\bibinfo{person}{Giorgos Giannakakis}, \bibinfo{person}{Matthew Pediaditis}, \bibinfo{person}{Dimitris Manousos}, \bibinfo{person}{Eleni Kazantzaki}, \bibinfo{person}{Franco Chiarugi}, \bibinfo{person}{Panagiotis~G. Simos}, \bibinfo{person}{Kostas Marias}, {and} \bibinfo{person}{Manolis Tsiknakis}.} \bibinfo{year}{2017}\natexlab{}.
\newblock \showarticletitle{Stress and anxiety detection using facial cues from videos}.
\newblock \bibinfo{journal}{\emph{Biomed. Signal Process. Control.}}  \bibinfo{volume}{31} (\bibinfo{year}{2017}), \bibinfo{pages}{89--101}.
\newblock


\bibitem[Gjoreski et~al\mbox{.}(2016)]%
        {Gjoreski2016ContinuousSD}
\bibfield{author}{\bibinfo{person}{Martin Gjoreski}, \bibinfo{person}{Hristijan Gjoreski}, \bibinfo{person}{Mitja Luvstrek}, {and} \bibinfo{person}{Matjavz Gams}.} \bibinfo{year}{2016}\natexlab{}.
\newblock \showarticletitle{Continuous stress detection using a wrist device: in laboratory and real life}.
\newblock \bibinfo{journal}{\emph{Proceedings of the 2016 ACM International Joint Conference on Pervasive and Ubiquitous Computing: Adjunct}} (\bibinfo{year}{2016}).
\newblock


\bibitem[Gloster et~al\mbox{.}(2008)]%
        {Gloster2008PsychometricPO}
\bibfield{author}{\bibinfo{person}{Andrew~T. Gloster}, \bibinfo{person}{Howard~M. Rhoades}, \bibinfo{person}{Diane~M. Novy}, \bibinfo{person}{Jens Klotsche}, \bibinfo{person}{Ashley~C Senior}, \bibinfo{person}{Mark~E. Kunik}, \bibinfo{person}{Nancy~L. Wilson}, {and} \bibinfo{person}{Melinda~A. Stanley}.} \bibinfo{year}{2008}\natexlab{}.
\newblock \showarticletitle{Psychometric properties of the Depression Anxiety and Stress Scale-21 in older primary care patients.}
\newblock \bibinfo{journal}{\emph{Journal of affective disorders}}  \bibinfo{volume}{110 3} (\bibinfo{year}{2008}), \bibinfo{pages}{248--59}.
\newblock


\bibitem[Golbeck et~al\mbox{.}(2011)]%
        {golbeck2011predicting}
\bibfield{author}{\bibinfo{person}{Jennifer Golbeck}, \bibinfo{person}{Cristina Robles}, \bibinfo{person}{Michon Edmondson}, {and} \bibinfo{person}{Karen Turner}.} \bibinfo{year}{2011}\natexlab{}.
\newblock \showarticletitle{Predicting personality from twitter}. In \bibinfo{booktitle}{\emph{2011 IEEE third international conference on privacy, security, risk and trust and 2011 IEEE third international conference on social computing}}. IEEE, \bibinfo{pages}{149--156}.
\newblock


\bibitem[Goyal and Fussell(2017)]%
        {Goyal2017}
\bibfield{author}{\bibinfo{person}{Nitesh Goyal} {and} \bibinfo{person}{Susan~R. Fussell}.} \bibinfo{year}{2017}\natexlab{}.
\newblock \showarticletitle{Intelligent Interruption Management Using Electro Dermal Activity Based Physiological Sensor for Collaborative Sensemaking}.
\newblock \bibinfo{journal}{\emph{Proc. ACM Interact. Mob. Wearable Ubiquitous Technol.}} \bibinfo{volume}{1}, \bibinfo{number}{3}, Article \bibinfo{articleno}{52} (\bibinfo{date}{sep} \bibinfo{year}{2017}), \bibinfo{numpages}{21}~pages.
\newblock
\urldef\tempurl%
\url{https://doi.org/10.1145/3130917}
\showDOI{\tempurl}


\bibitem[Guo et~al\mbox{.}(2022)]%
        {GuoMSLife}
\bibfield{author}{\bibinfo{person}{Gabriel Guo}, \bibinfo{person}{Hanbin Zhang}, \bibinfo{person}{Liuyi Yao}, \bibinfo{person}{Huining Li}, \bibinfo{person}{Chenhan Xu}, \bibinfo{person}{Zhengxiong Li}, {and} \bibinfo{person}{Wenyao Xu}.} \bibinfo{year}{2022}\natexlab{}.
\newblock \showarticletitle{MSLife: Digital Behavioral Phenotyping of Multiple Sclerosis Symptoms in the Wild Using Wearables and Graph-Based Statistical Analysis}.
\newblock \bibinfo{journal}{\emph{Proc. ACM Interact. Mob. Wearable Ubiquitous Technol.}} \bibinfo{volume}{5}, \bibinfo{number}{4}, Article \bibinfo{articleno}{158} (\bibinfo{date}{dec} \bibinfo{year}{2022}), \bibinfo{numpages}{35}~pages.
\newblock
\urldef\tempurl%
\url{https://doi.org/10.1145/3494970}
\showDOI{\tempurl}


\bibitem[Hellhammer et~al\mbox{.}(2010)]%
        {article}
\bibfield{author}{\bibinfo{person}{Dirk Hellhammer}, \bibinfo{person}{Arthur Stone}, \bibinfo{person}{Juliane Hellhammer}, {and} \bibinfo{person}{Joan Broderick}.} \bibinfo{year}{2010}\natexlab{}.
\newblock \showarticletitle{Measuring Stress}.
\newblock \bibinfo{journal}{\emph{Encyclopedia of behavioural neuroscience}} (\bibinfo{date}{12} \bibinfo{year}{2010}), \bibinfo{pages}{186--191}.
\newblock
\urldef\tempurl%
\url{https://doi.org/10.1016/B978-0-08-045396-5.00188-3}
\showDOI{\tempurl}


\bibitem[Hern{\'a}ndez et~al\mbox{.}(2014)]%
        {Hernndez2014UsingEA}
\bibfield{author}{\bibinfo{person}{Javier Hern{\'a}ndez}, \bibinfo{person}{Ivan Riobo}, \bibinfo{person}{Agata Rozga}, \bibinfo{person}{Gregory~D. Abowd}, {and} \bibinfo{person}{Rosalind~W. Picard}.} \bibinfo{year}{2014}\natexlab{}.
\newblock \showarticletitle{Using electrodermal activity to recognize ease of engagement in children during social interactions}.
\newblock \bibinfo{journal}{\emph{Proceedings of the 2014 ACM International Joint Conference on Pervasive and Ubiquitous Computing}} (\bibinfo{year}{2014}).
\newblock


\bibitem[Hovsepian et~al\mbox{.}(2015)]%
        {Hovsepian2015cStressTA}
\bibfield{author}{\bibinfo{person}{Karen Hovsepian}, \bibinfo{person}{Mustafa al’Absi}, \bibinfo{person}{Emre Ertin}, \bibinfo{person}{Tom~P Kamarck}, \bibinfo{person}{Motohiro Nakajima}, {and} \bibinfo{person}{Santosh Kumar}.} \bibinfo{year}{2015}\natexlab{}.
\newblock \showarticletitle{cStress: towards a gold standard for continuous stress assessment in the mobile environment}.
\newblock \bibinfo{journal}{\emph{Proceedings of the 2015 ACM International Joint Conference on Pervasive and Ubiquitous Computing}} (\bibinfo{year}{2015}).
\newblock


\bibitem[Huang et~al\mbox{.}(2016)]%
        {Huang2016AssessingSA}
\bibfield{author}{\bibinfo{person}{Yu Huang}, \bibinfo{person}{Haoyi Xiong}, \bibinfo{person}{Kevin Leach}, \bibinfo{person}{Yuyan Zhang}, \bibinfo{person}{Philip~I. Chow}, \bibinfo{person}{Karl~C. Fua}, \bibinfo{person}{Bethany~A. Teachman}, {and} \bibinfo{person}{Laura~E. Barnes}.} \bibinfo{year}{2016}\natexlab{}.
\newblock \showarticletitle{Assessing social anxiety using gps trajectories and point-of-interest data}.
\newblock \bibinfo{journal}{\emph{Proceedings of the 2016 ACM International Joint Conference on Pervasive and Ubiquitous Computing}} (\bibinfo{year}{2016}).
\newblock


\bibitem[Huppert and So(2011)]%
        {Huppert2011FlourishingAE}
\bibfield{author}{\bibinfo{person}{Felicia~A. Huppert} {and} \bibinfo{person}{Timothy T.~C. So}.} \bibinfo{year}{2011}\natexlab{}.
\newblock \showarticletitle{Flourishing Across Europe: Application of a New Conceptual Framework for Defining Well-Being}.
\newblock \bibinfo{journal}{\emph{Social Indicators Research}}  \bibinfo{volume}{110} (\bibinfo{year}{2011}), \bibinfo{pages}{837 -- 861}.
\newblock


\bibitem[Huynh et~al\mbox{.}(2018)]%
        {Huynh2018EngageMonME}
\bibfield{author}{\bibinfo{person}{Sinh Huynh}, \bibinfo{person}{Seungmin Kim}, \bibinfo{person}{Jeonggil Ko}, \bibinfo{person}{Rajesh~Krishna Balan}, {and} \bibinfo{person}{Youngki Lee}.} \bibinfo{year}{2018}\natexlab{}.
\newblock \showarticletitle{EngageMon: Multi-Modal Engagement Sensing for Mobile Games}.
\newblock \bibinfo{journal}{\emph{Proc. ACM Interact. Mob. Wearable Ubiquitous Technol.}}  \bibinfo{volume}{2} (\bibinfo{year}{2018}), \bibinfo{pages}{13:1--13:27}.
\newblock


\bibitem[Irazoki et~al\mbox{.}(2020)]%
        {irazoki2020technologies}
\bibfield{author}{\bibinfo{person}{Eider Irazoki}, \bibinfo{person}{Leslie~Mar{\'\i}a Contreras-Somoza}, \bibinfo{person}{Jos{\'e}~Miguel Toribio-Guzm{\'a}n}, \bibinfo{person}{Cristina Jenaro-R{\'\i}o}, \bibinfo{person}{Henri{\"e}Tte Van~der Roest}, {and} \bibinfo{person}{Manuel~A Franco-Mart{\'\i}n}.} \bibinfo{year}{2020}\natexlab{}.
\newblock \showarticletitle{Technologies for cognitive training and cognitive rehabilitation for people with mild cognitive impairment and dementia. A systematic review}.
\newblock \bibinfo{journal}{\emph{Frontiers in psychology}}  \bibinfo{volume}{11} (\bibinfo{year}{2020}), \bibinfo{pages}{648}.
\newblock


\bibitem[Jaimes et~al\mbox{.}(2007)]%
        {jaimes2007guest}
\bibfield{author}{\bibinfo{person}{Alejandro Jaimes}, \bibinfo{person}{Daniel Gatica-Perez}, \bibinfo{person}{Nicu Sebe}, {and} \bibinfo{person}{Thomas~S Huang}.} \bibinfo{year}{2007}\natexlab{}.
\newblock \showarticletitle{Guest Editors' Introduction: Human-Centered Computing--Toward a Human Revolution}.
\newblock \bibinfo{journal}{\emph{Computer}} \bibinfo{volume}{40}, \bibinfo{number}{5} (\bibinfo{year}{2007}), \bibinfo{pages}{30--34}.
\newblock


\bibitem[John and Srivastava(1999)]%
        {John1999TheBF}
\bibfield{author}{\bibinfo{person}{Oliver~P. John} {and} \bibinfo{person}{Sanjay Srivastava}.} \bibinfo{year}{1999}\natexlab{}.
\newblock \showarticletitle{The Big Five Trait taxonomy: History, measurement, and theoretical perspectives.}
\newblock


\bibitem[Kang(2021)]%
        {kang2021sample}
\bibfield{author}{\bibinfo{person}{Hyun Kang}.} \bibinfo{year}{2021}\natexlab{}.
\newblock \showarticletitle{Sample size determination and power analysis using the G* Power software}.
\newblock \bibinfo{journal}{\emph{Journal of educational evaluation for health professions}}  \bibinfo{volume}{18} (\bibinfo{year}{2021}).
\newblock


\bibitem[Kaur et~al\mbox{.}(2022)]%
        {kaur2022didn}
\bibfield{author}{\bibinfo{person}{Harmanpreet Kaur}, \bibinfo{person}{Daniel McDuff}, \bibinfo{person}{Alex~C Williams}, \bibinfo{person}{Jaime Teevan}, {and} \bibinfo{person}{Shamsi~T Iqbal}.} \bibinfo{year}{2022}\natexlab{}.
\newblock \showarticletitle{“I didn’t know I looked angry”: Characterizing observed emotion and reported affect at work}. In \bibinfo{booktitle}{\emph{Proceedings of the 2022 CHI Conference on Human Factors in Computing Systems}}. \bibinfo{pages}{1--18}.
\newblock


\bibitem[Kelley et~al\mbox{.}(2003)]%
        {kelley2003good}
\bibfield{author}{\bibinfo{person}{Kate Kelley}, \bibinfo{person}{Belinda Clark}, \bibinfo{person}{Vivienne Brown}, {and} \bibinfo{person}{John Sitzia}.} \bibinfo{year}{2003}\natexlab{}.
\newblock \showarticletitle{Good practice in the conduct and reporting of survey research}.
\newblock \bibinfo{journal}{\emph{International Journal for Quality in health care}} \bibinfo{volume}{15}, \bibinfo{number}{3} (\bibinfo{year}{2003}), \bibinfo{pages}{261--266}.
\newblock


\bibitem[Khwaja et~al\mbox{.}(2019)]%
        {khwaja}
\bibfield{author}{\bibinfo{person}{Mohammed Khwaja}, \bibinfo{person}{Sumer~S. Vaid}, \bibinfo{person}{Sara Zannone}, \bibinfo{person}{Gabriella~M. Harari}, \bibinfo{person}{A.~Aldo Faisal}, {and} \bibinfo{person}{Aleksandar Matic}.} \bibinfo{year}{2019}\natexlab{}.
\newblock \showarticletitle{Modeling Personality vs. Modeling Personalidad: In-the-Wild Mobile Data Analysis in Five Countries Suggests Cultural Impact on Personality Models}.
\newblock \bibinfo{journal}{\emph{Proc. ACM Interact. Mob. Wearable Ubiquitous Technol.}} \bibinfo{volume}{3}, \bibinfo{number}{3}, Article \bibinfo{articleno}{88} (\bibinfo{date}{sep} \bibinfo{year}{2019}), \bibinfo{numpages}{24}~pages.
\newblock
\urldef\tempurl%
\url{https://doi.org/10.1145/3351246}
\showDOI{\tempurl}


\bibitem[Kilkenny and Robinson(2018)]%
        {kilkenny2018data}
\bibfield{author}{\bibinfo{person}{Monique~F Kilkenny} {and} \bibinfo{person}{Kerin~M Robinson}.} \bibinfo{year}{2018}\natexlab{}.
\newblock \bibinfo{title}{Data quality:“Garbage in--garbage out”}.
\newblock , \bibinfo{numpages}{103--105}~pages.
\newblock


\bibitem[Kim et~al\mbox{.}(2019)]%
        {chatbot}
\bibfield{author}{\bibinfo{person}{Soomin Kim}, \bibinfo{person}{Joonhwan Lee}, {and} \bibinfo{person}{Gahgene Gweon}.} \bibinfo{year}{2019}\natexlab{}.
\newblock \showarticletitle{Comparing Data from Chatbot and Web Surveys: Effects of Platform and Conversational Style on Survey Response Quality}. In \bibinfo{booktitle}{\emph{Proceedings of the 2019 CHI Conference on Human Factors in Computing Systems}} (Glasgow, Scotland Uk) \emph{(\bibinfo{series}{CHI '19})}. \bibinfo{publisher}{Association for Computing Machinery}, \bibinfo{address}{New York, NY, USA}, \bibinfo{pages}{1–12}.
\newblock
\showISBNx{9781450359702}
\urldef\tempurl%
\url{https://doi.org/10.1145/3290605.3300316}
\showDOI{\tempurl}


\bibitem[King et~al\mbox{.}(2019)]%
        {microstress}
\bibfield{author}{\bibinfo{person}{Zachary~D. King}, \bibinfo{person}{Judith Moskowitz}, \bibinfo{person}{Begum Egilmez}, \bibinfo{person}{Shibo Zhang}, \bibinfo{person}{Lida Zhang}, \bibinfo{person}{Michael Bass}, \bibinfo{person}{John Rogers}, \bibinfo{person}{Roozbeh Ghaffari}, \bibinfo{person}{Laurie Wakschlag}, {and} \bibinfo{person}{Nabil Alshurafa}.} \bibinfo{year}{2019}\natexlab{}.
\newblock \showarticletitle{Micro-Stress EMA: A Passive Sensing Framework for Detecting in-the-Wild Stress in Pregnant Mothers}.
\newblock \bibinfo{journal}{\emph{Proc. ACM Interact. Mob. Wearable Ubiquitous Technol.}} \bibinfo{volume}{3}, \bibinfo{number}{3}, Article \bibinfo{articleno}{91} (\bibinfo{date}{sep} \bibinfo{year}{2019}), \bibinfo{numpages}{22}~pages.
\newblock
\urldef\tempurl%
\url{https://doi.org/10.1145/3351249}
\showDOI{\tempurl}


\bibitem[Kirschbaum et~al\mbox{.}(1993)]%
        {Kirschbaum1993TheS}
\bibfield{author}{\bibinfo{person}{Clemens Kirschbaum}, \bibinfo{person}{Karl~Martin Pirke}, {and} \bibinfo{person}{Dirk~H. Hellhammer}.} \bibinfo{year}{1993}\natexlab{}.
\newblock \showarticletitle{The 'Trier Social Stress Test'--a tool for investigating psychobiological stress responses in a laboratory setting.}
\newblock \bibinfo{journal}{\emph{Neuropsychobiology}}  \bibinfo{volume}{28 1-2} (\bibinfo{year}{1993}), \bibinfo{pages}{76--81}.
\newblock


\bibitem[Kosch et~al\mbox{.}(2018)]%
        {kosch2018look}
\bibfield{author}{\bibinfo{person}{Thomas Kosch}, \bibinfo{person}{Mariam Hassib}, \bibinfo{person}{Daniel Buschek}, {and} \bibinfo{person}{Albrecht Schmidt}.} \bibinfo{year}{2018}\natexlab{}.
\newblock \showarticletitle{Look into my eyes: using pupil dilation to estimate mental workload for task complexity adaptation}. In \bibinfo{booktitle}{\emph{Extended abstracts of the 2018 chi conference on human factors in computing systems}}. \bibinfo{pages}{1--6}.
\newblock


\bibitem[{Kroenke} et~al\mbox{.}(2009a)]%
        {kroenke2009phq4}
\bibfield{author}{\bibinfo{person}{Kurt {Kroenke}}, \bibinfo{person}{Robert~L. {Spitzer}}, \bibinfo{person}{Janet~B.W. {Williams}}, {and} \bibinfo{person}{Bernd {Löwe}}.} \bibinfo{year}{2009}\natexlab{a}.
\newblock \showarticletitle{An Ultra-brief Screening Scale for Anxiety and Depression: The PHQ-4}.
\newblock \bibinfo{journal}{\emph{Psychosomatics}} \bibinfo{volume}{50}, \bibinfo{number}{6} (\bibinfo{year}{2009}), \bibinfo{pages}{613--621}.
\newblock


\bibitem[Kroenke et~al\mbox{.}(2010)]%
        {KROENKE2010345}
\bibfield{author}{\bibinfo{person}{Kurt Kroenke}, \bibinfo{person}{Robert~L. Spitzer}, \bibinfo{person}{Janet~B.W. Williams}, {and} \bibinfo{person}{Bernd Löwe}.} \bibinfo{year}{2010}\natexlab{}.
\newblock \showarticletitle{The Patient Health Questionnaire Somatic, Anxiety, and Depressive Symptom Scales: a systematic review}.
\newblock \bibinfo{journal}{\emph{General Hospital Psychiatry}} \bibinfo{volume}{32}, \bibinfo{number}{4} (\bibinfo{year}{2010}), \bibinfo{pages}{345--359}.
\newblock
\showISSN{0163-8343}
\urldef\tempurl%
\url{https://doi.org/10.1016/j.genhosppsych.2010.03.006}
\showDOI{\tempurl}


\bibitem[{Kroenke} et~al\mbox{.}(2009b)]%
        {kroenke2009phq8}
\bibfield{author}{\bibinfo{person}{Kurt {Kroenke}}, \bibinfo{person}{Tara~W. {Strine}}, \bibinfo{person}{Robert~L. {Spitzer}}, \bibinfo{person}{Janet~B.W. {Williams}}, \bibinfo{person}{Joyce~T. {Berry}}, {and} \bibinfo{person}{Ali~H. {Mokdad}}.} \bibinfo{year}{2009}\natexlab{b}.
\newblock \showarticletitle{The PHQ-8 as a Measure of Current Depression in the General Population}.
\newblock \bibinfo{journal}{\emph{Journal of Affective Disorders}} \bibinfo{volume}{114}, \bibinfo{number}{1} (\bibinfo{year}{2009}), \bibinfo{pages}{163--173}.
\newblock


\bibitem[Krupp et~al\mbox{.}(1989)]%
        {Krupp1989TheFS}
\bibfield{author}{\bibinfo{person}{Lauren~B. Krupp}, \bibinfo{person}{Nicholas~G. Larocca}, \bibinfo{person}{Joanne Muir-Nash}, {and} \bibinfo{person}{Alfred~D. Steinberg}.} \bibinfo{year}{1989}\natexlab{}.
\newblock \showarticletitle{The fatigue severity scale. Application to patients with multiple sclerosis and systemic lupus erythematosus.}
\newblock \bibinfo{journal}{\emph{Archives of neurology}}  \bibinfo{volume}{46 10} (\bibinfo{year}{1989}), \bibinfo{pages}{1121--3}.
\newblock


\bibitem[Larradet et~al\mbox{.}(2019)]%
        {Larradet}
\bibfield{author}{\bibinfo{person}{Fanny Larradet}, \bibinfo{person}{Radoslaw Niewiadomski}, \bibinfo{person}{Giacinto Barresi}, {and} \bibinfo{person}{Leonardo~S. Mattos}.} \bibinfo{year}{2019}\natexlab{}.
\newblock \showarticletitle{Appraisal Theory-Based Mobile App for Physiological Data Collection and Labelling in the Wild}. In \bibinfo{booktitle}{\emph{Adjunct Proceedings of the 2019 ACM International Joint Conference on Pervasive and Ubiquitous Computing and Proceedings of the 2019 ACM International Symposium on Wearable Computers}} (London, United Kingdom) \emph{(\bibinfo{series}{UbiComp/ISWC '19 Adjunct})}. \bibinfo{publisher}{Association for Computing Machinery}, \bibinfo{address}{New York, NY, USA}, \bibinfo{pages}{752–756}.
\newblock
\showISBNx{9781450368698}
\urldef\tempurl%
\url{https://doi.org/10.1145/3341162.3345595}
\showDOI{\tempurl}


\bibitem[Lascio et~al\mbox{.}(2020)]%
        {Multi-Sensor}
\bibfield{author}{\bibinfo{person}{Elena~Di Lascio}, \bibinfo{person}{Shkurta Gashi}, \bibinfo{person}{Juan~Sebastian Hidalgo}, \bibinfo{person}{Beatrice Nale}, \bibinfo{person}{Maike~E. Debus}, {and} \bibinfo{person}{Silvia Santini}.} \bibinfo{year}{2020}\natexlab{}.
\newblock \showarticletitle{A Multi-Sensor Approach to Automatically Recognize Breaks and Work Activities of Knowledge Workers in Academia}.
\newblock \bibinfo{journal}{\emph{Proceedings of the ACM on Interactive, Mobile, Wearable and Ubiquitous Technologies}}  \bibinfo{volume}{4} (\bibinfo{year}{2020}), \bibinfo{pages}{1 -- 20}.
\newblock


\bibitem[Lascio et~al\mbox{.}(2018)]%
        {Lascio2018UnobtrusiveAO}
\bibfield{author}{\bibinfo{person}{Elena~Di Lascio}, \bibinfo{person}{Shkurta Gashi}, {and} \bibinfo{person}{Silvia Santini}.} \bibinfo{year}{2018}\natexlab{}.
\newblock \showarticletitle{Unobtrusive Assessment of Students' Emotional Engagement during Lectures Using Electrodermal Activity Sensors}.
\newblock \bibinfo{journal}{\emph{Proceedings of the ACM on Interactive, Mobile, Wearable and Ubiquitous Technologies}}  \bibinfo{volume}{2} (\bibinfo{year}{2018}), \bibinfo{pages}{1 -- 21}.
\newblock


\bibitem[Lee et~al\mbox{.}(2016)]%
        {Lee2016OSNMT}
\bibfield{author}{\bibinfo{person}{James~Alexander Lee}, \bibinfo{person}{Christos Efstratiou}, {and} \bibinfo{person}{Lu Bai}.} \bibinfo{year}{2016}\natexlab{}.
\newblock \showarticletitle{OSN mood tracking: exploring the use of online social network activity as an indicator of mood changes}.
\newblock \bibinfo{journal}{\emph{Proceedings of the 2016 ACM International Joint Conference on Pervasive and Ubiquitous Computing: Adjunct}} (\bibinfo{year}{2016}).
\newblock


\bibitem[Lewinsohn et~al\mbox{.}(1997)]%
        {Lewinsohn1997CenterFE}
\bibfield{author}{\bibinfo{person}{Peter~M. Lewinsohn}, \bibinfo{person}{John~R. Seeley}, \bibinfo{person}{Robert~Edmund Roberts}, {and} \bibinfo{person}{Nicholas~B. Allen}.} \bibinfo{year}{1997}\natexlab{}.
\newblock \showarticletitle{Center for Epidemiologic Studies Depression Scale (CES-D) as a screening instrument for depression among community-residing older adults.}
\newblock \bibinfo{journal}{\emph{Psychology and aging}}  \bibinfo{volume}{12 2} (\bibinfo{year}{1997}), \bibinfo{pages}{277--87}.
\newblock


\bibitem[Li and Sano(2020)]%
        {Li2020ExtractionAI}
\bibfield{author}{\bibinfo{person}{Boning Li} {and} \bibinfo{person}{Akane Sano}.} \bibinfo{year}{2020}\natexlab{}.
\newblock \showarticletitle{Extraction and Interpretation of Deep Autoencoder-based Temporal Features from Wearables for Forecasting Personalized Mood, Health, and Stress}.
\newblock \bibinfo{journal}{\emph{Proceedings of the ACM on Interactive, Mobile, Wearable and Ubiquitous Technologies}}  \bibinfo{volume}{4} (\bibinfo{year}{2020}), \bibinfo{pages}{1 -- 26}.
\newblock


\bibitem[Li et~al\mbox{.}(2016)]%
        {Li2016EustressOD}
\bibfield{author}{\bibinfo{person}{Chun-Tung Li}, \bibinfo{person}{Jiannong Cao}, {and} \bibinfo{person}{Tim M.~H. Li}.} \bibinfo{year}{2016}\natexlab{}.
\newblock \showarticletitle{Eustress or distress: an empirical study of perceived stress in everyday college life}.
\newblock \bibinfo{journal}{\emph{Proceedings of the 2016 ACM International Joint Conference on Pervasive and Ubiquitous Computing: Adjunct}} (\bibinfo{year}{2016}).
\newblock


\bibitem[Liu and Wronski(2018)]%
        {liu2018trap}
\bibfield{author}{\bibinfo{person}{Mingnan Liu} {and} \bibinfo{person}{Laura Wronski}.} \bibinfo{year}{2018}\natexlab{}.
\newblock \showarticletitle{Trap questions in online surveys: Results from three web survey experiments}.
\newblock \bibinfo{journal}{\emph{International Journal of Market Research}} \bibinfo{volume}{60}, \bibinfo{number}{1} (\bibinfo{year}{2018}), \bibinfo{pages}{32--49}.
\newblock


\bibitem[Lockhart et~al\mbox{.}(2012)]%
        {lockhart2012applications}
\bibfield{author}{\bibinfo{person}{Jeffrey~W Lockhart}, \bibinfo{person}{Tony Pulickal}, {and} \bibinfo{person}{Gary~M Weiss}.} \bibinfo{year}{2012}\natexlab{}.
\newblock \showarticletitle{Applications of mobile activity recognition}. In \bibinfo{booktitle}{\emph{Proceedings of the 2012 ACM conference on ubiquitous computing}}. \bibinfo{pages}{1054--1058}.
\newblock


\bibitem[Lupton(2014)]%
        {lupton2014self}
\bibfield{author}{\bibinfo{person}{Deborah Lupton}.} \bibinfo{year}{2014}\natexlab{}.
\newblock \showarticletitle{Self-tracking cultures: towards a sociology of personal informatics}. In \bibinfo{booktitle}{\emph{Proceedings of the 26th Australian computer-human interaction conference on designing futures: The future of design}}. \bibinfo{pages}{77--86}.
\newblock


\bibitem[Luvstrek and Kaluvza(2009)]%
        {luvstrek2009fall}
\bibfield{author}{\bibinfo{person}{Mitja Luvstrek} {and} \bibinfo{person}{Bovstjan Kaluvza}.} \bibinfo{year}{2009}\natexlab{}.
\newblock \showarticletitle{Fall detection and activity recognition with machine learning}.
\newblock \bibinfo{journal}{\emph{Informatica}} \bibinfo{volume}{33}, \bibinfo{number}{2} (\bibinfo{year}{2009}).
\newblock


\bibitem[Luxton et~al\mbox{.}(2011)]%
        {luxton2011mhealth}
\bibfield{author}{\bibinfo{person}{David~D Luxton}, \bibinfo{person}{Russell~A McCann}, \bibinfo{person}{Nigel~E Bush}, \bibinfo{person}{Matthew~C Mishkind}, {and} \bibinfo{person}{Greg~M Reger}.} \bibinfo{year}{2011}\natexlab{}.
\newblock \showarticletitle{mHealth for mental health: Integrating smartphone technology in behavioral healthcare.}
\newblock \bibinfo{journal}{\emph{Professional Psychology: Research and Practice}} \bibinfo{volume}{42}, \bibinfo{number}{6} (\bibinfo{year}{2011}), \bibinfo{pages}{505}.
\newblock


\bibitem[Malhotra(2008)]%
        {malhotra2008completion}
\bibfield{author}{\bibinfo{person}{Neil Malhotra}.} \bibinfo{year}{2008}\natexlab{}.
\newblock \showarticletitle{Completion Time and Response Order Effects in Web Surveys}.
\newblock \bibinfo{journal}{\emph{Public Opinion Quarterly}} \bibinfo{volume}{72}, \bibinfo{number}{5} (\bibinfo{year}{2008}), \bibinfo{pages}{914--934}.
\newblock


\bibitem[Mar{\^o}co et~al\mbox{.}(2016)]%
        {Marco2016UniversitySE}
\bibfield{author}{\bibinfo{person}{Jo{\~a}o Mar{\^o}co}, \bibinfo{person}{Ana~L{\'u}cia Mar{\^o}co}, \bibinfo{person}{Juliana Alvares Duarte~Bonini Campos}, {and} \bibinfo{person}{Jennifer Fredricks}.} \bibinfo{year}{2016}\natexlab{}.
\newblock \showarticletitle{University student’s engagement: development of the University Student Engagement Inventory (USEI)}.
\newblock \bibinfo{journal}{\emph{Psicologia: Reflex{\~a}o e Cr{\'i}tica}}  \bibinfo{volume}{29} (\bibinfo{year}{2016}).
\newblock


\bibitem[Mattick and Clarke(1998)]%
        {MATTICK1998455}
\bibfield{author}{\bibinfo{person}{Richard~P. Mattick} {and} \bibinfo{person}{J.Christopher Clarke}.} \bibinfo{year}{1998}\natexlab{}.
\newblock \showarticletitle{Development and validation of measures of social phobia scrutiny fear and social interaction anxiety11Editor’s note: This article was written before the development of some contemporary measures of social phobia, such as the Social Phobia and Anxiety Inventory (Turner et al., 1989). We have invited this article for publication because of the growing interest in the scales described therein. S.T.}
\newblock \bibinfo{journal}{\emph{Behaviour Research and Therapy}} \bibinfo{volume}{36}, \bibinfo{number}{4} (\bibinfo{year}{1998}), \bibinfo{pages}{455--470}.
\newblock
\showISSN{0005-7967}
\urldef\tempurl%
\url{https://doi.org/10.1016/S0005-7967(97)10031-6}
\showDOI{\tempurl}


\bibitem[Medhat et~al\mbox{.}(2014)]%
        {medhat2014sentiment}
\bibfield{author}{\bibinfo{person}{Walaa Medhat}, \bibinfo{person}{Ahmed Hassan}, {and} \bibinfo{person}{Hoda Korashy}.} \bibinfo{year}{2014}\natexlab{}.
\newblock \showarticletitle{Sentiment analysis algorithms and applications: A survey}.
\newblock \bibinfo{journal}{\emph{Ain Shams engineering journal}} \bibinfo{volume}{5}, \bibinfo{number}{4} (\bibinfo{year}{2014}), \bibinfo{pages}{1093--1113}.
\newblock


\bibitem[Meegahapola et~al\mbox{.}(2023)]%
        {Generalization}
\bibfield{author}{\bibinfo{person}{Lakmal Meegahapola}, \bibinfo{person}{William Droz}, \bibinfo{person}{Peter Kun}, \bibinfo{person}{Amalia de G\"{o}tzen}, \bibinfo{person}{Chaitanya Nutakki}, \bibinfo{person}{Shyam Diwakar}, \bibinfo{person}{Salvador~Ruiz Correa}, \bibinfo{person}{Donglei Song}, \bibinfo{person}{Hao Xu}, \bibinfo{person}{Miriam Bidoglia}, \bibinfo{person}{George Gaskell}, \bibinfo{person}{Altangerel Chagnaa}, \bibinfo{person}{Amarsanaa Ganbold}, \bibinfo{person}{Tsolmon Zundui}, \bibinfo{person}{Carlo Caprini}, \bibinfo{person}{Daniele Miorandi}, \bibinfo{person}{Alethia Hume}, \bibinfo{person}{Jose~Luis Zarza}, \bibinfo{person}{Luca Cernuzzi}, \bibinfo{person}{Ivano Bison}, \bibinfo{person}{Marcelo~Rodas Britez}, \bibinfo{person}{Matteo Busso}, \bibinfo{person}{Ronald Chenu-Abente}, \bibinfo{person}{Can G\"{u}nel}, \bibinfo{person}{Fausto Giunchiglia}, \bibinfo{person}{Laura Schelenz}, {and} \bibinfo{person}{Daniel Gatica-Perez}.} \bibinfo{year}{2023}\natexlab{}.
\newblock \showarticletitle{Generalization and Personalization of Mobile Sensing-Based Mood Inference Models: An Analysis of College Students in Eight Countries}.
\newblock \bibinfo{journal}{\emph{Proc. ACM Interact. Mob. Wearable Ubiquitous Technol.}} \bibinfo{volume}{6}, \bibinfo{number}{4}, Article \bibinfo{articleno}{176} (\bibinfo{date}{jan} \bibinfo{year}{2023}), \bibinfo{numpages}{32}~pages.
\newblock
\urldef\tempurl%
\url{https://doi.org/10.1145/3569483}
\showDOI{\tempurl}


\bibitem[Miller and Horst(2020)]%
        {miller2020digital}
\bibfield{author}{\bibinfo{person}{Daniel Miller} {and} \bibinfo{person}{Heather~A Horst}.} \bibinfo{year}{2020}\natexlab{}.
\newblock \showarticletitle{The digital and the human: A prospectus for digital anthropology}.
\newblock In \bibinfo{booktitle}{\emph{Digital anthropology}}. \bibinfo{publisher}{Routledge}, \bibinfo{pages}{3--35}.
\newblock


\bibitem[Mirjafari et~al\mbox{.}(2019)]%
        {Mirjafari2019DifferentiatingHA}
\bibfield{author}{\bibinfo{person}{Shayan Mirjafari}, \bibinfo{person}{Kizito Masaba}, \bibinfo{person}{Ted Grover}, \bibinfo{person}{Weichen Wang}, \bibinfo{person}{Pino~G. Audia}, \bibinfo{person}{Andrew~T. Campbell}, \bibinfo{person}{N. Chawla}, \bibinfo{person}{Vedant~Das Swain}, \bibinfo{person}{Munmun~De Choudhury}, \bibinfo{person}{Anind~K. Dey}, \bibinfo{person}{Sidney~K. D’Mello}, \bibinfo{person}{Ge Gao}, \bibinfo{person}{Julie~M. Gregg}, \bibinfo{person}{Krithika Jagannath}, \bibinfo{person}{Kaifeng Jiang}, \bibinfo{person}{Suwen Lin}, \bibinfo{person}{Qiang Liu}, \bibinfo{person}{Gloria Mark}, \bibinfo{person}{Gonzalo~J. Mart{\'i}nez}, \bibinfo{person}{Stephen~M. Mattingly}, \bibinfo{person}{Edward Moskal}, \bibinfo{person}{Raghu Mulukutla}, \bibinfo{person}{Subigya Nepal}, \bibinfo{person}{Kari~A. Nies}, \bibinfo{person}{Manikanta~D. Reddy}, \bibinfo{person}{Pablo Robles-Granda}, \bibinfo{person}{Koustuv Saha}, \bibinfo{person}{Anusha Sirigiri}, {and} \bibinfo{person}{Aaron~D. Striegel}.}
  \bibinfo{year}{2019}\natexlab{}.
\newblock \showarticletitle{Differentiating Higher and Lower Job Performers in the Workplace Using Mobile Sensing}.
\newblock \bibinfo{journal}{\emph{Proceedings of the ACM on Interactive, Mobile, Wearable and Ubiquitous Technologies}}  \bibinfo{volume}{3} (\bibinfo{year}{2019}), \bibinfo{pages}{1 -- 24}.
\newblock


\bibitem[Mishra et~al\mbox{.}(2018)]%
        {Mishra2018TheCF}
\bibfield{author}{\bibinfo{person}{Varun Mishra}, \bibinfo{person}{Gunnar Pope}, \bibinfo{person}{Sarah~E. Lord}, \bibinfo{person}{Stephanie Lewia}, \bibinfo{person}{Byron~M. Lowens}, \bibinfo{person}{Kelly~E. Caine}, \bibinfo{person}{Sougata Sen}, \bibinfo{person}{Ryan~J. Halter}, {and} \bibinfo{person}{David~F. Kotz}.} \bibinfo{year}{2018}\natexlab{}.
\newblock \showarticletitle{The Case for a Commodity Hardware Solution for Stress Detection}.
\newblock \bibinfo{journal}{\emph{Proceedings of the 2018 ACM International Joint Conference and 2018 International Symposium on Pervasive and Ubiquitous Computing and Wearable Computers}} (\bibinfo{year}{2018}).
\newblock


\bibitem[Mozos et~al\mbox{.}(2017)]%
        {Mozos2017StressDU}
\bibfield{author}{\bibinfo{person}{{\'O}scar~Mart{\'i}nez Mozos}, \bibinfo{person}{Virginia Sandulescu}, \bibinfo{person}{Sally Andrews}, \bibinfo{person}{David Ellis}, \bibinfo{person}{Nicola Bellotto}, \bibinfo{person}{Radu Dobrescu}, {and} \bibinfo{person}{J.~M. Ferr{\'a}ndez}.} \bibinfo{year}{2017}\natexlab{}.
\newblock \showarticletitle{Stress Detection Using Wearable Physiological and Sociometric Sensors}.
\newblock \bibinfo{journal}{\emph{International journal of neural systems}}  \bibinfo{volume}{27 2} (\bibinfo{year}{2017}), \bibinfo{pages}{1650041}.
\newblock


\bibitem[Mutepfa and Tapera(2019)]%
        {Mutepfa2019}
\bibfield{author}{\bibinfo{person}{Magen~Mhaka Mutepfa} {and} \bibinfo{person}{Roy Tapera}.} \bibinfo{year}{2019}\natexlab{}.
\newblock \bibinfo{booktitle}{\emph{Traditional Survey and Questionnaire Platforms}}.
\newblock \bibinfo{publisher}{Springer Singapore}, \bibinfo{address}{Singapore}, \bibinfo{pages}{541--558}.
\newblock
\showISBNx{978-981-10-5251-4}
\urldef\tempurl%
\url{https://doi.org/10.1007/978-981-10-5251-4_89}
\showDOI{\tempurl}


\bibitem[Naim et~al\mbox{.}(2015)]%
        {Naim}
\bibfield{author}{\bibinfo{person}{Iftekhar Naim}, \bibinfo{person}{M.~Iftekhar Tanveer}, \bibinfo{person}{Daniel Gildea}, {and} \bibinfo{person}{Mohammed~Ehsan Hoque}.} \bibinfo{year}{2015}\natexlab{}.
\newblock \showarticletitle{Automated prediction and analysis of job interview performance: The role of what you say and how you say it}. In \bibinfo{booktitle}{\emph{2015 11th IEEE International Conference and Workshops on Automatic Face and Gesture Recognition (FG)}}, Vol.~\bibinfo{volume}{1}. \bibinfo{pages}{1--6}.
\newblock
\urldef\tempurl%
\url{https://doi.org/10.1109/FG.2015.7163127}
\showDOI{\tempurl}


\bibitem[Nepal et~al\mbox{.}(2020)]%
        {Nepal2020DetectingJP}
\bibfield{author}{\bibinfo{person}{Subigya Nepal}, \bibinfo{person}{Shayan Mirjafari}, \bibinfo{person}{Gonzalo~J. Mart{\'i}nez}, \bibinfo{person}{Pino~G. Audia}, \bibinfo{person}{Aaron~D. Striegel}, {and} \bibinfo{person}{Andrew~T. Campbell}.} \bibinfo{year}{2020}\natexlab{}.
\newblock \showarticletitle{Detecting Job Promotion in Information Workers Using Mobile Sensing}.
\newblock \bibinfo{journal}{\emph{Proceedings of the ACM on Interactive, Mobile, Wearable and Ubiquitous Technologies}}  \bibinfo{volume}{4} (\bibinfo{year}{2020}), \bibinfo{pages}{1 -- 28}.
\newblock


\bibitem[Panda et~al\mbox{.}(2020)]%
        {panda2020audio}
\bibfield{author}{\bibinfo{person}{Renato Panda}, \bibinfo{person}{Ricardo~Manuel Malheiro}, {and} \bibinfo{person}{Rui~Pedro Paiva}.} \bibinfo{year}{2020}\natexlab{}.
\newblock \showarticletitle{Audio features for music emotion recognition: a survey}.
\newblock \bibinfo{journal}{\emph{IEEE Transactions on Affective Computing}} (\bibinfo{year}{2020}).
\newblock


\bibitem[Paulhus et~al\mbox{.}(2007)]%
        {paulhus2007self}
\bibfield{author}{\bibinfo{person}{Delroy~L Paulhus}, \bibinfo{person}{Simine Vazire}, {et~al\mbox{.}}} \bibinfo{year}{2007}\natexlab{}.
\newblock \showarticletitle{The self-report method}.
\newblock \bibinfo{journal}{\emph{Handbook of research methods in personality psychology}} \bibinfo{volume}{1}, \bibinfo{number}{2007} (\bibinfo{year}{2007}), \bibinfo{pages}{224--239}.
\newblock


\bibitem[Peterson(2022)]%
        {peterson_2022}
\bibfield{author}{\bibinfo{person}{Ashley~L. Peterson}.} \bibinfo{year}{2022}\natexlab{}.
\newblock \bibinfo{title}{What is... psychological testing}.
\newblock
\newblock
\urldef\tempurl%
\url{https://mentalhealthathome.org/2020/09/04/what-is-psychological-testing/}
\showURL{%
\tempurl}


\bibitem[Pollak et~al\mbox{.}(2011)]%
        {pollak2011pam}
\bibfield{author}{\bibinfo{person}{John~P Pollak}, \bibinfo{person}{Phil Adams}, {and} \bibinfo{person}{Geri Gay}.} \bibinfo{year}{2011}\natexlab{}.
\newblock \showarticletitle{PAM: a photographic affect meter for frequent, in situ measurement of affect}. In \bibinfo{booktitle}{\emph{Proceedings of the SIGCHI conference on Human factors in computing systems}}. \bibinfo{pages}{725--734}.
\newblock


\bibitem[Pratt et~al\mbox{.}(2004)]%
        {pratt2004incorporating}
\bibfield{author}{\bibinfo{person}{Wanda Pratt}, \bibinfo{person}{Madhu~C Reddy}, \bibinfo{person}{David~W McDonald}, \bibinfo{person}{Peter Tarczy-Hornoch}, {and} \bibinfo{person}{John~H Gennari}.} \bibinfo{year}{2004}\natexlab{}.
\newblock \showarticletitle{Incorporating ideas from computer-supported cooperative work}.
\newblock \bibinfo{journal}{\emph{Journal of biomedical informatics}} \bibinfo{volume}{37}, \bibinfo{number}{2} (\bibinfo{year}{2004}), \bibinfo{pages}{128--137}.
\newblock


\bibitem[Rehman et~al\mbox{.}(2020)]%
        {Rehman2020DepressionAA}
\bibfield{author}{\bibinfo{person}{Usama Rehman}, \bibinfo{person}{Mohammad~Ghazi Shahnawaz}, \bibinfo{person}{Neda~Haseeb Khan}, \bibinfo{person}{Korsi~Dorene Kharshiing}, \bibinfo{person}{Masrat Khursheed}, \bibinfo{person}{Kaveri Gupta}, \bibinfo{person}{Drishti Kashyap}, {and} \bibinfo{person}{Ritika Uniyal}.} \bibinfo{year}{2020}\natexlab{}.
\newblock \showarticletitle{Depression, Anxiety and Stress Among Indians in Times of Covid-19 Lockdown}.
\newblock \bibinfo{journal}{\emph{Community Mental Health Journal}}  \bibinfo{volume}{57} (\bibinfo{year}{2020}), \bibinfo{pages}{42 -- 48}.
\newblock


\bibitem[Rodrigues et~al\mbox{.}(2015)]%
        {Rodrigues}
\bibfield{author}{\bibinfo{person}{João G.~P. Rodrigues}, \bibinfo{person}{Mariana Kaiseler}, \bibinfo{person}{Ana Aguiar}, \bibinfo{person}{João~P. Silva~Cunha}, {and} \bibinfo{person}{João Barros}.} \bibinfo{year}{2015}\natexlab{}.
\newblock \showarticletitle{A Mobile Sensing Approach to Stress Detection and Memory Activation for Public Bus Drivers}.
\newblock \bibinfo{journal}{\emph{IEEE Transactions on Intelligent Transportation Systems}} \bibinfo{volume}{16}, \bibinfo{number}{6} (\bibinfo{year}{2015}), \bibinfo{pages}{3294--3303}.
\newblock
\urldef\tempurl%
\url{https://doi.org/10.1109/TITS.2015.2445314}
\showDOI{\tempurl}


\bibitem[Rubin et~al\mbox{.}(2015)]%
        {Towards}
\bibfield{author}{\bibinfo{person}{Jonathan Rubin}, \bibinfo{person}{Hoda Eldardiry}, \bibinfo{person}{Rui Abreu}, \bibinfo{person}{Shane Ahern}, \bibinfo{person}{Honglu Du}, \bibinfo{person}{Ashish Pattekar}, {and} \bibinfo{person}{Daniel~G. Bobrow}.} \bibinfo{year}{2015}\natexlab{}.
\newblock \showarticletitle{Towards a Mobile and Wearable System for Predicting Panic Attacks}. In \bibinfo{booktitle}{\emph{Proceedings of the 2015 ACM International Joint Conference on Pervasive and Ubiquitous Computing}} (Osaka, Japan) \emph{(\bibinfo{series}{UbiComp '15})}. \bibinfo{publisher}{Association for Computing Machinery}, \bibinfo{address}{New York, NY, USA}, \bibinfo{pages}{529–533}.
\newblock
\showISBNx{9781450335744}
\urldef\tempurl%
\url{https://doi.org/10.1145/2750858.2805834}
\showDOI{\tempurl}


\bibitem[Russell(1996)]%
        {ucla}
\bibfield{author}{\bibinfo{person}{Daniel Russell}.} \bibinfo{year}{1996}\natexlab{}.
\newblock \showarticletitle{UCLA Loneliness Scale (Version 3): Reliability, Validity, and Factor Structure}.
\newblock \bibinfo{journal}{\emph{Journal of personality assessment}}  \bibinfo{volume}{66} (\bibinfo{date}{03} \bibinfo{year}{1996}), \bibinfo{pages}{20--40}.
\newblock
\urldef\tempurl%
\url{https://doi.org/10.1207/s15327752jpa6601_2}
\showDOI{\tempurl}


\bibitem[Saha et~al\mbox{.}(2017)]%
        {Saha2017InferringMI}
\bibfield{author}{\bibinfo{person}{Koustuv Saha}, \bibinfo{person}{Larry Chan}, \bibinfo{person}{Kaya de Barbaro}, \bibinfo{person}{Gregory~D. Abowd}, {and} \bibinfo{person}{Munmun~De Choudhury}.} \bibinfo{year}{2017}\natexlab{}.
\newblock \showarticletitle{Inferring Mood Instability on Social Media by Leveraging Ecological Momentary Assessments}.
\newblock \bibinfo{journal}{\emph{Proceedings of the ACM on Interactive, Mobile, Wearable and Ubiquitous Technologies}}  \bibinfo{volume}{1} (\bibinfo{year}{2017}), \bibinfo{pages}{1 -- 27}.
\newblock


\bibitem[Saha et~al\mbox{.}(2021)]%
        {Saha}
\bibfield{author}{\bibinfo{person}{Koustuv Saha}, \bibinfo{person}{Ted Grover}, \bibinfo{person}{Stephen Mattingly}, \bibinfo{person}{Vedant Das~Swain}, \bibinfo{person}{Pranshu Gupta}, \bibinfo{person}{Gonzalo Martinez}, \bibinfo{person}{Pablo Robles-Granda}, \bibinfo{person}{Gloria Mark}, \bibinfo{person}{Aaron Striegel}, {and} \bibinfo{person}{Munmun Choudhury}.} \bibinfo{year}{2021}\natexlab{}.
\newblock \showarticletitle{Person-Centered Predictions of Psychological Constructs with Social Media Contextualized by Multimodal Sensing}.
\newblock \bibinfo{journal}{\emph{Proceedings of the ACM on Interactive Mobile Wearable and Ubiquitous Technologies}}  \bibinfo{volume}{5} (\bibinfo{date}{03} \bibinfo{year}{2021}), \bibinfo{pages}{32}.
\newblock
\urldef\tempurl%
\url{https://doi.org/10.1145/3448117}
\showDOI{\tempurl}


\bibitem[Samyoun et~al\mbox{.}(2022)]%
        {Samyoun}
\bibfield{author}{\bibinfo{person}{Sirat Samyoun}, \bibinfo{person}{Md~Mofijul Islam}, \bibinfo{person}{Tariq Iqbal}, {and} \bibinfo{person}{John Stankovic}.} \bibinfo{year}{2022}\natexlab{}.
\newblock \showarticletitle{M3Sense: Affect-Agnostic Multitask Representation Learning Using Multimodal Wearable Sensors}.
\newblock \bibinfo{journal}{\emph{Proceedings of the ACM on Interactive, Mobile, Wearable and Ubiquitous Technologies}}  \bibinfo{volume}{6} (\bibinfo{date}{07} \bibinfo{year}{2022}), \bibinfo{pages}{1--32}.
\newblock
\urldef\tempurl%
\url{https://doi.org/10.1145/3534600}
\showDOI{\tempurl}


\bibitem[Sharma* and Mansotra(2019)]%
        {Sharma*_Mansotra_2019}
\bibfield{author}{\bibinfo{person}{Dr.~Archana Sharma*} {and} \bibinfo{person}{Dr.~Vibhakar Mansotra}.} \bibinfo{year}{2019}\natexlab{}.
\newblock \bibinfo{title}{Multimodal Decision-level Group Sentiment Prediction of Students in Classrooms}.
\newblock , \bibinfo{numpages}{4902–4909}~pages.
\newblock
\urldef\tempurl%
\url{https://doi.org/10.35940/ijitee.l3549.1081219}
\showDOI{\tempurl}


\bibitem[Shear et~al\mbox{.}(2001)]%
        {SHEAR2001293}
\bibfield{author}{\bibinfo{person}{M.Katherine Shear}, \bibinfo{person}{Paola Rucci}, \bibinfo{person}{Jenna Williams}, \bibinfo{person}{Ellen Frank}, \bibinfo{person}{Victoria Grochocinski}, \bibinfo{person}{Joni {Vander Bilt}}, \bibinfo{person}{Patricia Houck}, {and} \bibinfo{person}{Tracey Wang}.} \bibinfo{year}{2001}\natexlab{}.
\newblock \showarticletitle{Reliability and validity of the Panic Disorder Severity Scale: replication and extension}.
\newblock \bibinfo{journal}{\emph{Journal of Psychiatric Research}} \bibinfo{volume}{35}, \bibinfo{number}{5} (\bibinfo{year}{2001}), \bibinfo{pages}{293--296}.
\newblock
\showISSN{0022-3956}
\urldef\tempurl%
\url{https://doi.org/10.1016/S0022-3956(01)00028-0}
\showDOI{\tempurl}


\bibitem[Sinha et~al\mbox{.}(2010)]%
        {sinha2010human}
\bibfield{author}{\bibinfo{person}{Gaurav Sinha}, \bibinfo{person}{Rahul Shahi}, {and} \bibinfo{person}{Mani Shankar}.} \bibinfo{year}{2010}\natexlab{}.
\newblock \showarticletitle{Human computer interaction}. In \bibinfo{booktitle}{\emph{2010 3rd International Conference on Emerging Trends in Engineering and Technology}}. IEEE, \bibinfo{pages}{1--4}.
\newblock


\bibitem[Smith et~al\mbox{.}(2019)]%
        {incspr}
\bibfield{author}{\bibinfo{person}{Michael Smith}, \bibinfo{person}{Maryam Witte}, \bibinfo{person}{Sarah Rocha}, {and} \bibinfo{person}{Mathias Basner}.} \bibinfo{year}{2019}\natexlab{}.
\newblock \showarticletitle{Effectiveness of incentives and follow-up on increasing survey response rates and participation in field studies}.
\newblock \bibinfo{journal}{\emph{BMC Medical Research Methodology}}  \bibinfo{volume}{19} (\bibinfo{date}{12} \bibinfo{year}{2019}), \bibinfo{pages}{230}.
\newblock
\urldef\tempurl%
\url{https://doi.org/10.1186/s12874-019-0868-8}
\showDOI{\tempurl}


\bibitem[Solso et~al\mbox{.}(2005)]%
        {solso2005cognitive}
\bibfield{author}{\bibinfo{person}{Robert~L Solso}, \bibinfo{person}{M~Kimberly MacLin}, {and} \bibinfo{person}{Otto~H MacLin}.} \bibinfo{year}{2005}\natexlab{}.
\newblock \bibinfo{booktitle}{\emph{Cognitive psychology}}.
\newblock \bibinfo{publisher}{Pearson Education New Zealand}.
\newblock


\bibitem[Sonnenberg et~al\mbox{.}(2012)]%
        {SONNENBERG20121037}
\bibfield{author}{\bibinfo{person}{Bettina Sonnenberg}, \bibinfo{person}{Michaela Riediger}, \bibinfo{person}{Cornelia Wrzus}, {and} \bibinfo{person}{Gert~G. Wagner}.} \bibinfo{year}{2012}\natexlab{}.
\newblock \showarticletitle{Measuring time use in surveys – Concordance of survey and experience sampling measures}.
\newblock \bibinfo{journal}{\emph{Social Science Research}} \bibinfo{volume}{41}, \bibinfo{number}{5} (\bibinfo{year}{2012}), \bibinfo{pages}{1037--1052}.
\newblock
\showISSN{0049-089X}
\urldef\tempurl%
\url{https://doi.org/10.1016/j.ssresearch.2012.03.013}
\showDOI{\tempurl}


\bibitem[Spielberger et~al\mbox{.}(1970)]%
        {Spielberger1970ManualFT}
\bibfield{author}{\bibinfo{person}{Charles~Donald Spielberger}, \bibinfo{person}{Richard~L. Gorsuch}, {and} \bibinfo{person}{Robert~E. Lushene}.} \bibinfo{year}{1970}\natexlab{}.
\newblock \showarticletitle{Manual for the State-Trait Anxiety Inventory}.
\newblock


\bibitem[Spitzer et~al\mbox{.}(2006)]%
        {spitzer2006brief}
\bibfield{author}{\bibinfo{person}{Robert~L Spitzer}, \bibinfo{person}{Kurt Kroenke}, \bibinfo{person}{Janet~BW Williams}, {and} \bibinfo{person}{Bernd L{\"o}we}.} \bibinfo{year}{2006}\natexlab{}.
\newblock \showarticletitle{A brief measure for assessing generalized anxiety disorder: the GAD-7}.
\newblock \bibinfo{journal}{\emph{Archives of internal medicine}} \bibinfo{volume}{166}, \bibinfo{number}{10} (\bibinfo{year}{2006}), \bibinfo{pages}{1092--1097}.
\newblock


\bibitem[Steimer(2002)]%
        {Steimer2002TheBO}
\bibfield{author}{\bibinfo{person}{Th. Steimer}.} \bibinfo{year}{2002}\natexlab{}.
\newblock \showarticletitle{The biology of fear- and anxiety-related behaviors}.
\newblock \bibinfo{journal}{\emph{Dialogues in Clinical Neuroscience}}  \bibinfo{volume}{4} (\bibinfo{year}{2002}), \bibinfo{pages}{231 -- 249}.
\newblock


\bibitem[Stone et~al\mbox{.}(1991)]%
        {stone1991measuring}
\bibfield{author}{\bibinfo{person}{Arthur~A Stone}, \bibinfo{person}{Ronald~C Kessler}, {and} \bibinfo{person}{Jennifer~A Haythomthwatte}.} \bibinfo{year}{1991}\natexlab{}.
\newblock \showarticletitle{Measuring daily events and experiences: Decisions for the researcher}.
\newblock \bibinfo{journal}{\emph{Journal of personality}} \bibinfo{volume}{59}, \bibinfo{number}{3} (\bibinfo{year}{1991}), \bibinfo{pages}{575--607}.
\newblock


\bibitem[Tag et~al\mbox{.}(2022)]%
        {tag2022emotion}
\bibfield{author}{\bibinfo{person}{Benjamin Tag}, \bibinfo{person}{Zhanna Sarsenbayeva}, \bibinfo{person}{Anna~L Cox}, \bibinfo{person}{Greg Wadley}, \bibinfo{person}{Jorge Goncalves}, {and} \bibinfo{person}{Vassilis Kostakos}.} \bibinfo{year}{2022}\natexlab{}.
\newblock \showarticletitle{Emotion trajectories in smartphone use: Towards recognizing emotion regulation in-the-wild}.
\newblock \bibinfo{journal}{\emph{International Journal of Human-Computer Studies}}  \bibinfo{volume}{166} (\bibinfo{year}{2022}), \bibinfo{pages}{102872}.
\newblock


\bibitem[Taigman et~al\mbox{.}(2014)]%
        {taigman2014deepface}
\bibfield{author}{\bibinfo{person}{Yaniv Taigman}, \bibinfo{person}{Ming Yang}, \bibinfo{person}{Marc'Aurelio Ranzato}, {and} \bibinfo{person}{Lior Wolf}.} \bibinfo{year}{2014}\natexlab{}.
\newblock \showarticletitle{Deepface: Closing the gap to human-level performance in face verification}. In \bibinfo{booktitle}{\emph{Proceedings of the IEEE conference on computer vision and pattern recognition}}. \bibinfo{pages}{1701--1708}.
\newblock


\bibitem[Taylor et~al\mbox{.}(2015)]%
        {Taylor2015UsingPS}
\bibfield{author}{\bibinfo{person}{Brandon~T. Taylor}, \bibinfo{person}{Anind~K. Dey}, \bibinfo{person}{Daniel~P. Siewiorek}, {and} \bibinfo{person}{Asim Smailagic}.} \bibinfo{year}{2015}\natexlab{}.
\newblock \showarticletitle{Using physiological sensors to detect levels of user frustration induced by system delays}.
\newblock \bibinfo{journal}{\emph{Proceedings of the 2015 ACM International Joint Conference on Pervasive and Ubiquitous Computing}} (\bibinfo{year}{2015}).
\newblock


\bibitem[Terzimehi{\'c} et~al\mbox{.}(2023)]%
        {terzimehic2023tale}
\bibfield{author}{\bibinfo{person}{Na{\dj}a Terzimehi{\'c}}, \bibinfo{person}{Sarah Aragon-Hahner}, {and} \bibinfo{person}{Heinrich Hussmann}.} \bibinfo{year}{2023}\natexlab{}.
\newblock \showarticletitle{The Tale of a Complicated Relationship: Insights from Users' Love/Breakup Letters to Their Smartphones before and during the COVID-19 Pandemic}.
\newblock \bibinfo{journal}{\emph{Proceedings of the ACM on Interactive, Mobile, Wearable and Ubiquitous Technologies}} \bibinfo{volume}{7}, \bibinfo{number}{1} (\bibinfo{year}{2023}), \bibinfo{pages}{1--34}.
\newblock


\bibitem[Tlachac et~al\mbox{.}(2022a)]%
        {Tlachac}
\bibfield{author}{\bibinfo{person}{ML Tlachac}, \bibinfo{person}{Ricardo Flores}, \bibinfo{person}{Miranda Reisch}, \bibinfo{person}{Katie Houskeeper}, {and} \bibinfo{person}{Elke Rundensteiner}.} \bibinfo{year}{2022}\natexlab{a}.
\newblock \showarticletitle{DepreST-CAT: Retrospective Smartphone Call and Text Logs Collected during the COVID-19 Pandemic to Screen for Mental Illnesses}.
\newblock \bibinfo{journal}{\emph{Proceedings of the ACM on Interactive, Mobile, Wearable and Ubiquitous Technologies}}  \bibinfo{volume}{6} (\bibinfo{date}{07} \bibinfo{year}{2022}), \bibinfo{pages}{1--32}.
\newblock
\urldef\tempurl%
\url{https://doi.org/10.1145/3534596}
\showDOI{\tempurl}


\bibitem[Tlachac et~al\mbox{.}(2022b)]%
        {Tlachac2022}
\bibfield{author}{\bibinfo{person}{ML Tlachac}, \bibinfo{person}{Ricardo Flores}, \bibinfo{person}{Miranda Reisch}, \bibinfo{person}{Rimsha Kayastha}, \bibinfo{person}{Nina Taurich}, \bibinfo{person}{Veronica Melican}, \bibinfo{person}{Connor Bruneau}, \bibinfo{person}{Hunter Caouette}, \bibinfo{person}{Joshua Lovering}, \bibinfo{person}{Ermal Toto}, {and} \bibinfo{person}{Elke Rundensteiner}.} \bibinfo{year}{2022}\natexlab{b}.
\newblock \showarticletitle{StudentSADD: Rapid Mobile Depression and Suicidal Ideation Screening of College Students during the Coronavirus Pandemic}.
\newblock \bibinfo{journal}{\emph{Proceedings of the ACM on Interactive, Mobile, Wearable and Ubiquitous Technologies}}  \bibinfo{volume}{6} (\bibinfo{date}{07} \bibinfo{year}{2022}), \bibinfo{pages}{1--32}.
\newblock
\urldef\tempurl%
\url{https://doi.org/10.1145/3534604}
\showDOI{\tempurl}


\bibitem[Tong et~al\mbox{.}(2019)]%
        {Tong2019TrackingFA}
\bibfield{author}{\bibinfo{person}{C. Tong}, \bibinfo{person}{Matthew~J. Craner}, \bibinfo{person}{Matthieu Vegreville}, {and} \bibinfo{person}{Nicholas~D. Lane}.} \bibinfo{year}{2019}\natexlab{}.
\newblock \showarticletitle{Tracking Fatigue and Health State in Multiple Sclerosis Patients Using Connnected Wellness Devices}.
\newblock \bibinfo{journal}{\emph{Proceedings of the ACM on Interactive, Mobile, Wearable and Ubiquitous Technologies}}  \bibinfo{volume}{3} (\bibinfo{year}{2019}), \bibinfo{pages}{1 -- 19}.
\newblock


\bibitem[Trull and Ebner-Priemer(2009)]%
        {Trull2009UsingES}
\bibfield{author}{\bibinfo{person}{Timothy~J. Trull} {and} \bibinfo{person}{Ulrich~W. Ebner-Priemer}.} \bibinfo{year}{2009}\natexlab{}.
\newblock \showarticletitle{Using experience sampling methods/ecological momentary assessment (ESM/EMA) in clinical assessment and clinical research: introduction to the special section.}
\newblock \bibinfo{journal}{\emph{Psychological assessment}}  \bibinfo{volume}{21 4} (\bibinfo{year}{2009}), \bibinfo{pages}{457--62}.
\newblock


\bibitem[Van~Berkel et~al\mbox{.}(2017)]%
        {van2017gamification}
\bibfield{author}{\bibinfo{person}{Niels Van~Berkel}, \bibinfo{person}{Jorge Goncalves}, \bibinfo{person}{Simo Hosio}, {and} \bibinfo{person}{Vassilis Kostakos}.} \bibinfo{year}{2017}\natexlab{}.
\newblock \showarticletitle{Gamification of mobile experience sampling improves data quality and quantity}.
\newblock \bibinfo{journal}{\emph{Proceedings of the ACM on Interactive, Mobile, Wearable and Ubiquitous Technologies}} \bibinfo{volume}{1}, \bibinfo{number}{3} (\bibinfo{year}{2017}), \bibinfo{pages}{1--21}.
\newblock


\bibitem[Wallace et~al\mbox{.}(2018)]%
        {Wallace2018TheCS}
\bibfield{author}{\bibinfo{person}{Meredith~L. Wallace}, \bibinfo{person}{Molly~H. Carter}, {and} \bibinfo{person}{Satish Iyengar}.} \bibinfo{year}{2018}\natexlab{}.
\newblock \showarticletitle{The Current State of EMA and ESM Study Design in Mood Disorders Research: A Comprehensive Summary and Analysis}.
\newblock


\bibitem[Wang et~al\mbox{.}(2016)]%
        {Wang2016CrossCheckTP}
\bibfield{author}{\bibinfo{person}{Rui Wang}, \bibinfo{person}{M.~S.~Hane Aung}, \bibinfo{person}{Saeed Abdullah}, \bibinfo{person}{Rachel~M Brian}, \bibinfo{person}{Andrew~T. Campbell}, \bibinfo{person}{Tanzeem Choudhury}, \bibinfo{person}{Marta Hauser}, \bibinfo{person}{John~M. Kane}, \bibinfo{person}{Michael Merrill}, \bibinfo{person}{Emily~A. Scherer}, \bibinfo{person}{Vincent Wen-Sheng Tseng}, {and} \bibinfo{person}{Dror Ben-Zeev}.} \bibinfo{year}{2016}\natexlab{}.
\newblock \showarticletitle{CrossCheck: toward passive sensing and detection of mental health changes in people with schizophrenia}.
\newblock \bibinfo{journal}{\emph{Proceedings of the 2016 ACM International Joint Conference on Pervasive and Ubiquitous Computing}} (\bibinfo{year}{2016}).
\newblock


\bibitem[Wang et~al\mbox{.}(2014)]%
        {studentlife}
\bibfield{author}{\bibinfo{person}{Rui Wang}, \bibinfo{person}{Fanglin Chen}, \bibinfo{person}{Zhenyu Chen}, \bibinfo{person}{Tianxing Li}, \bibinfo{person}{Gabriella Harari}, \bibinfo{person}{Stefanie Tignor}, \bibinfo{person}{Xia Zhou}, \bibinfo{person}{Dror Ben-Zeev}, {and} \bibinfo{person}{Andrew~T. Campbell}.} \bibinfo{year}{2014}\natexlab{}.
\newblock \showarticletitle{StudentLife: Assessing Mental Health, Academic Performance and Behavioral Trends of College Students Using Smartphones}. In \bibinfo{booktitle}{\emph{Proceedings of the 2014 ACM International Joint Conference on Pervasive and Ubiquitous Computing}} (Seattle, Washington) \emph{(\bibinfo{series}{UbiComp '14})}. \bibinfo{publisher}{Association for Computing Machinery}, \bibinfo{address}{New York, NY, USA}, \bibinfo{pages}{3–14}.
\newblock
\showISBNx{9781450329682}
\urldef\tempurl%
\url{https://doi.org/10.1145/2632048.2632054}
\showURL{%
\tempurl}


\bibitem[Wang et~al\mbox{.}(2015)]%
        {Wang2015SmartGPAHS}
\bibfield{author}{\bibinfo{person}{Rui Wang}, \bibinfo{person}{Gabriella~M. Harari}, \bibinfo{person}{Peilin Hao}, \bibinfo{person}{Xia Zhou}, {and} \bibinfo{person}{Andrew~T. Campbell}.} \bibinfo{year}{2015}\natexlab{}.
\newblock \showarticletitle{SmartGPA: how smartphones can assess and predict academic performance of college students}.
\newblock \bibinfo{journal}{\emph{Proceedings of the 2015 ACM International Joint Conference on Pervasive and Ubiquitous Computing}} (\bibinfo{year}{2015}).
\newblock


\bibitem[Wang et~al\mbox{.}(2017)]%
        {Wang2017PredictingST}
\bibfield{author}{\bibinfo{person}{Rui Wang}, \bibinfo{person}{Weichen Wang}, \bibinfo{person}{M.~S.~Hane Aung}, \bibinfo{person}{Dror Ben-Zeev}, \bibinfo{person}{Rachel~M Brian}, \bibinfo{person}{Andrew~T. Campbell}, \bibinfo{person}{Tanzeem Choudhury}, \bibinfo{person}{Marta Hauser}, \bibinfo{person}{John~M. Kane}, \bibinfo{person}{Emily~A. Scherer}, {and} \bibinfo{person}{Megan Walsh}.} \bibinfo{year}{2017}\natexlab{}.
\newblock \showarticletitle{Predicting Symptom Trajectories of Schizophrenia using Mobile Sensing}.
\newblock \bibinfo{journal}{\emph{Proceedings of the ACM on Interactive, Mobile, Wearable and Ubiquitous Technologies}}  \bibinfo{volume}{1} (\bibinfo{year}{2017}), \bibinfo{pages}{1 -- 24}.
\newblock


\bibitem[Wang et~al\mbox{.}(2018b)]%
        {wang2018trackingdepression}
\bibfield{author}{\bibinfo{person}{Rui Wang}, \bibinfo{person}{Weichen Wang}, \bibinfo{person}{Alex DaSilva}, \bibinfo{person}{Jeremy~F Huckins}, \bibinfo{person}{William~M Kelley}, \bibinfo{person}{Todd~F Heatherton}, {and} \bibinfo{person}{Andrew~T Campbell}.} \bibinfo{year}{2018}\natexlab{b}.
\newblock \showarticletitle{Tracking Depression Dynamics in College Students using Mobile Phone and Wearable Sensing}.
\newblock \bibinfo{journal}{\emph{Proceedings of the ACM on Interactive, Mobile, Wearable and Ubiquitous Technologies}} \bibinfo{volume}{2}, \bibinfo{number}{1} (\bibinfo{year}{2018}), \bibinfo{pages}{1--26}.
\newblock


\bibitem[Wang et~al\mbox{.}(2018c)]%
        {Wang2018TrackingDD}
\bibfield{author}{\bibinfo{person}{Rui Wang}, \bibinfo{person}{Weichen Wang}, \bibinfo{person}{Alex~W DaSilva}, \bibinfo{person}{Jeremy~F. Huckins}, \bibinfo{person}{William~M. Kelley}, \bibinfo{person}{Todd~F. Heatherton}, {and} \bibinfo{person}{Andrew~T. Campbell}.} \bibinfo{year}{2018}\natexlab{c}.
\newblock \showarticletitle{Tracking Depression Dynamics in College Students Using Mobile Phone and Wearable Sensing}.
\newblock \bibinfo{journal}{\emph{Proceedings of the ACM on Interactive, Mobile, Wearable and Ubiquitous Technologies}}  \bibinfo{volume}{2} (\bibinfo{year}{2018}), \bibinfo{pages}{1 -- 26}.
\newblock


\bibitem[Wang et~al\mbox{.}(2018a)]%
        {Wang2018SensingBC}
\bibfield{author}{\bibinfo{person}{Weichen Wang}, \bibinfo{person}{Gabriella~M. Harari}, \bibinfo{person}{Rui Wang}, \bibinfo{person}{Sandrine~R. M{\"u}ller}, \bibinfo{person}{Shayan Mirjafari}, \bibinfo{person}{Kizito Masaba}, {and} \bibinfo{person}{Andrew~T. Campbell}.} \bibinfo{year}{2018}\natexlab{a}.
\newblock \showarticletitle{Sensing Behavioral Change over Time}.
\newblock \bibinfo{journal}{\emph{Proceedings of the ACM on Interactive, Mobile, Wearable and Ubiquitous Technologies}}  \bibinfo{volume}{2} (\bibinfo{year}{2018}), \bibinfo{pages}{1 -- 21}.
\newblock


\bibitem[Wang et~al\mbox{.}(2022)]%
        {WangFirstGen}
\bibfield{author}{\bibinfo{person}{Weichen Wang}, \bibinfo{person}{Subigya Nepal}, \bibinfo{person}{Jeremy~F. Huckins}, \bibinfo{person}{Lessley Hernandez}, \bibinfo{person}{Vlado Vojdanovski}, \bibinfo{person}{Dante Mack}, \bibinfo{person}{Jane Plomp}, \bibinfo{person}{Arvind Pillai}, \bibinfo{person}{Mikio Obuchi}, \bibinfo{person}{Alex daSilva}, \bibinfo{person}{Eilis Murphy}, \bibinfo{person}{Elin Hedlund}, \bibinfo{person}{Courtney Rogers}, \bibinfo{person}{Meghan Meyer}, {and} \bibinfo{person}{Andrew Campbell}.} \bibinfo{year}{2022}\natexlab{}.
\newblock \showarticletitle{First-Gen Lens: Assessing Mental Health of First-Generation Students across Their First Year at College Using Mobile Sensing}.
\newblock \bibinfo{journal}{\emph{Proc. ACM Interact. Mob. Wearable Ubiquitous Technol.}} \bibinfo{volume}{6}, \bibinfo{number}{2}, Article \bibinfo{articleno}{95} (\bibinfo{date}{jul} \bibinfo{year}{2022}), \bibinfo{numpages}{32}~pages.
\newblock
\urldef\tempurl%
\url{https://doi.org/10.1145/3543194}
\showDOI{\tempurl}


\bibitem[Wash et~al\mbox{.}(2017)]%
        {wash2017can}
\bibfield{author}{\bibinfo{person}{Rick Wash}, \bibinfo{person}{Emilee Rader}, {and} \bibinfo{person}{Chris Fennell}.} \bibinfo{year}{2017}\natexlab{}.
\newblock \showarticletitle{Can People Self-Report Security Accurately? Agreement Between Self-Report and Behavioral Measures}. In \bibinfo{booktitle}{\emph{Proceedings of the 2017 CHI Conference on Human Factors in Computing Systems}}. \bibinfo{pages}{2228--2232}.
\newblock


\bibitem[Wilson et~al\mbox{.}(2021)]%
        {Wilson2021}
\bibfield{author}{\bibinfo{person}{Justin~C. Wilson}, \bibinfo{person}{Suku Nair}, \bibinfo{person}{Sandro Scielzo}, {and} \bibinfo{person}{Eric~C. Larson}.} \bibinfo{year}{2021}\natexlab{}.
\newblock \showarticletitle{Objective Measures of Cognitive Load Using Deep Multi-Modal Learning: A Use-Case in Aviation}.
\newblock \bibinfo{journal}{\emph{Proc. ACM Interact. Mob. Wearable Ubiquitous Technol.}} \bibinfo{volume}{5}, \bibinfo{number}{1}, Article \bibinfo{articleno}{40} (\bibinfo{date}{mar} \bibinfo{year}{2021}), \bibinfo{numpages}{35}~pages.
\newblock
\urldef\tempurl%
\url{https://doi.org/10.1145/3448111}
\showDOI{\tempurl}


\bibitem[Xu et~al\mbox{.}(2019)]%
        {Xu2019LeveragingRB}
\bibfield{author}{\bibinfo{person}{Xuhai Xu}, \bibinfo{person}{Prerna Chikersal}, \bibinfo{person}{Afsaneh Doryab}, \bibinfo{person}{Daniella~K. Villalba}, \bibinfo{person}{Janine~M. Dutcher}, \bibinfo{person}{Michael~J. Tumminia}, \bibinfo{person}{Tim Althoff}, \bibinfo{person}{Sheldon Cohen}, \bibinfo{person}{Kasey~G. Creswell}, \bibinfo{person}{J.~David Creswell}, \bibinfo{person}{Jennifer Mankoff}, {and} \bibinfo{person}{Anind~K. Dey}.} \bibinfo{year}{2019}\natexlab{}.
\newblock \showarticletitle{Leveraging Routine Behavior and Contextually-Filtered Features for Depression Detection among College Students}.
\newblock \bibinfo{journal}{\emph{Proceedings of the ACM on Interactive, Mobile, Wearable and Ubiquitous Technologies}}  \bibinfo{volume}{3} (\bibinfo{year}{2019}), \bibinfo{pages}{1 -- 33}.
\newblock


\bibitem[Xu et~al\mbox{.}(2021)]%
        {levereging2021}
\bibfield{author}{\bibinfo{person}{Xuhai Xu}, \bibinfo{person}{Prerna Chikersal}, \bibinfo{person}{Janine~M. Dutcher}, \bibinfo{person}{Yasaman~S. Sefidgar}, \bibinfo{person}{Woosuk Seo}, \bibinfo{person}{Michael~J. Tumminia}, \bibinfo{person}{Daniella~K. Villalba}, \bibinfo{person}{Sheldon Cohen}, \bibinfo{person}{Kasey~G. Creswell}, \bibinfo{person}{J.~David Creswell}, \bibinfo{person}{Afsaneh Doryab}, \bibinfo{person}{Paula~S. Nurius}, \bibinfo{person}{Eve Riskin}, \bibinfo{person}{Anind~K. Dey}, {and} \bibinfo{person}{Jennifer Mankoff}.} \bibinfo{year}{2021}\natexlab{}.
\newblock \showarticletitle{Leveraging Collaborative-Filtering for Personalized Behavior Modeling: A Case Study of Depression Detection among College Students}.
\newblock \bibinfo{journal}{\emph{Proc. ACM Interact. Mob. Wearable Ubiquitous Technol.}} \bibinfo{volume}{5}, \bibinfo{number}{1}, Article \bibinfo{articleno}{41} (\bibinfo{date}{mar} \bibinfo{year}{2021}), \bibinfo{numpages}{27}~pages.
\newblock
\urldef\tempurl%
\url{https://doi.org/10.1145/3448107}
\showDOI{\tempurl}


\bibitem[Xu et~al\mbox{.}(2023)]%
        {GLOBEM}
\bibfield{author}{\bibinfo{person}{Xuhai Xu}, \bibinfo{person}{Xin Liu}, \bibinfo{person}{Han Zhang}, \bibinfo{person}{Weichen Wang}, \bibinfo{person}{Subigya Nepal}, \bibinfo{person}{Yasaman Sefidgar}, \bibinfo{person}{Woosuk Seo}, \bibinfo{person}{Kevin~S. Kuehn}, \bibinfo{person}{Jeremy~F. Huckins}, \bibinfo{person}{Margaret~E. Morris}, \bibinfo{person}{Paula~S. Nurius}, \bibinfo{person}{Eve~A. Riskin}, \bibinfo{person}{Shwetak Patel}, \bibinfo{person}{Tim Althoff}, \bibinfo{person}{Andrew Campbell}, \bibinfo{person}{Anind~K. Dey}, {and} \bibinfo{person}{Jennifer Mankoff}.} \bibinfo{year}{2023}\natexlab{}.
\newblock \showarticletitle{GLOBEM: Cross-Dataset Generalization of Longitudinal Human Behavior Modeling}.
\newblock \bibinfo{journal}{\emph{Proc. ACM Interact. Mob. Wearable Ubiquitous Technol.}} \bibinfo{volume}{6}, \bibinfo{number}{4}, Article \bibinfo{articleno}{190} (\bibinfo{date}{jan} \bibinfo{year}{2023}), \bibinfo{numpages}{34}~pages.
\newblock
\urldef\tempurl%
\url{https://doi.org/10.1145/3569485}
\showDOI{\tempurl}


\bibitem[Yu and Sano(2023)]%
        {Yu}
\bibfield{author}{\bibinfo{person}{Han Yu} {and} \bibinfo{person}{Akane Sano}.} \bibinfo{year}{2023}\natexlab{}.
\newblock \showarticletitle{Semi-Supervised Learning for Wearable-Based Momentary Stress Detection in the Wild}.
\newblock \bibinfo{journal}{\emph{Proc. ACM Interact. Mob. Wearable Ubiquitous Technol.}} \bibinfo{volume}{7}, \bibinfo{number}{2}, Article \bibinfo{articleno}{80} (\bibinfo{date}{jun} \bibinfo{year}{2023}), \bibinfo{numpages}{23}~pages.
\newblock
\urldef\tempurl%
\url{https://doi.org/10.1145/3596246}
\showDOI{\tempurl}


\bibitem[Yu et~al\mbox{.}(2017)]%
        {sprinc}
\bibfield{author}{\bibinfo{person}{Shengchao Yu}, \bibinfo{person}{Howard~E Alper}, \bibinfo{person}{Angela-Maithy Nguyen}, \bibinfo{person}{Robert~M. Brackbill}, \bibinfo{person}{Lennon Turner}, \bibinfo{person}{Deborah~J. Walker}, \bibinfo{person}{Carey~B. Maslow}, {and} \bibinfo{person}{Kimberly~Caramanica Zweig}.} \bibinfo{year}{2017}\natexlab{}.
\newblock \showarticletitle{The effectiveness of a monetary incentive offer on survey response rates and response completeness in a longitudinal study}.
\newblock \bibinfo{journal}{\emph{BMC Medical Research Methodology}}  \bibinfo{volume}{17} (\bibinfo{year}{2017}).
\newblock


\bibitem[Zeng et~al\mbox{.}(2007)]%
        {zeng2007survey}
\bibfield{author}{\bibinfo{person}{Zhihong Zeng}, \bibinfo{person}{Maja Pantic}, \bibinfo{person}{Glenn~I Roisman}, {and} \bibinfo{person}{Thomas~S Huang}.} \bibinfo{year}{2007}\natexlab{}.
\newblock \showarticletitle{A survey of affect recognition methods: audio, visual and spontaneous expressions}. In \bibinfo{booktitle}{\emph{Proceedings of the 9th international conference on Multimodal interfaces}}. \bibinfo{pages}{126--133}.
\newblock


\bibitem[Zhang et~al\mbox{.}(2018a)]%
        {Moodexplorer}
\bibfield{author}{\bibinfo{person}{Xiao Zhang}, \bibinfo{person}{Wenzhong Li}, \bibinfo{person}{Xu Chen}, {and} \bibinfo{person}{Sanglu Lu}.} \bibinfo{year}{2018}\natexlab{a}.
\newblock \showarticletitle{Moodexplorer: Towards Compound Emotion Detection via Smartphone Sensing}.
\newblock \bibinfo{journal}{\emph{Proceedings of the ACM on Interactive, Mobile, Wearable and Ubiquitous Technologies}} \bibinfo{volume}{1}, \bibinfo{number}{4} (\bibinfo{year}{2018}), \bibinfo{pages}{1--30}.
\newblock


\bibitem[Zhang et~al\mbox{.}(2018b)]%
        {Zhang2018MoodExplorerTC}
\bibfield{author}{\bibinfo{person}{X. Zhang}, \bibinfo{person}{Wenzhong Li}, \bibinfo{person}{Xu Chen}, {and} \bibinfo{person}{Sanglu Lu}.} \bibinfo{year}{2018}\natexlab{b}.
\newblock \showarticletitle{MoodExplorer: Towards Compound Emotion Detection via Smartphone Sensing}.
\newblock \bibinfo{journal}{\emph{Proc. ACM Interact. Mob. Wearable Ubiquitous Technol.}}  \bibinfo{volume}{1} (\bibinfo{year}{2018}), \bibinfo{pages}{176:1--176:30}.
\newblock


\bibitem[Zhang et~al\mbox{.}(2018c)]%
        {Zhang2018TeamSenseAP}
\bibfield{author}{\bibinfo{person}{Yanxia Zhang}, \bibinfo{person}{Jeffrey Olenick}, \bibinfo{person}{Chu Hsiang~Daisy Chang}, \bibinfo{person}{Steve W.~J. Kozlowski}, {and} \bibinfo{person}{H. Hung}.} \bibinfo{year}{2018}\natexlab{c}.
\newblock \showarticletitle{TeamSense: Assessing Personal Affect and Group Cohesion in Small Teams through Dyadic Interaction and Behavior Analysis with Wearable Sensors}.
\newblock \bibinfo{journal}{\emph{Proc. ACM Interact. Mob. Wearable Ubiquitous Technol.}}  \bibinfo{volume}{2} (\bibinfo{year}{2018}), \bibinfo{pages}{150:1--150:22}.
\newblock


\end{thebibliography}


\end{document}